\documentclass[aps,pre,reprint,superscriptaddress]{revtex4-2}
\usepackage{graphicx,color}
\usepackage[T1]{fontenc}       
\usepackage{bm}
\begin{document}

\title{
Electrification of water interface
}
\author{Yuki Uematsu}
\address{$^1$ Department of Physics, Kyushu University, Fukuoka 819-0395, Japan}
\date{\today}
\begin{abstract}
The surface charge of a water interface determines many fundamental processes in physical chemistry and interface science, and it has been intensively studied for over a hundred years. 
We summarize experimental methods to characterize the surface charge densities developed so far: electrokinetics, double-layer force measurements, potentiometric titration, surface-sensitive nonlinear spectroscopy, and surface-sensitive mass spectrometry.
Then, we elucidate physical ion adsorption and chemical electrification as examples of electrification mechanisms. 
In the end, novel effects on surface electrification are discussed in detail. 
We believe that this clear overview of state of the art in a charged water interface will surely help the fundamental progress of physics and chemistry at interfaces in the future.
\end{abstract}

\maketitle

\section{Introduction}

The surface charge of water interfaces determines many physical, chemical, biological, and engineering processes, such as electrokinetics \cite{Dukhin1974,Joly2020,Swan2020,Nikita2021}, electric double layer capacitors \cite{Ben2018,Rene2020}, electrochemical reactions \cite{Fenn2000,Mayrhofer2011,Dunwell2018}, colloid and protein stability in solutions \cite{Russel1989,Adachi1995,Kobayashi2005,Nadine2016}, surface forces \cite{IsraelachviliBook,Israelachvili2010}, biological and artificial membrane transport \cite{Naranjo1994,Miedema2002,SiriaPoncharalBianceFulcrandBlasePurcellBocquet2013,Alice2020}, and food science \cite{Milk2020}. 
Although studies on the water’s surface charge and the electric double layer have a long history \cite{Biscombe2017,Lyklema2003,Wall2010}, they are still intensely debated in the fields of physics and chemistry \cite{Zhang2021, Barisic2021, Uematsu2020, Bonn2020, Bonn2020_2, Allen2021, Mishra2020,Carpenter2020, Roke2020, Alice2020, Grassian2019,Hassanali2020}.

Electro-osmosis was first reported by Reuss in 1807 \cite{Biscombe2017}, where water flows through a plug of clay by applying an external voltage.
Half a century later, Helmholtz derived a quantitative equation of the electro-osmotic mobility related to the surface potential \cite{Wall2010, Lyklema2003}. 
At the beginning of the 20th century, the mean-field description of electrostatics and hydrodynamics of charged water interfaces was developed using the Boltzmann distribution of ions \cite{Gouy1910, Chapman1913, DebyeHuckel1923}.
This enables quantitative characterization of the surface charge density by the electro-osmotic mobility.
To extract the surface charge density of colloids, electrophoresis theories have been developed \cite{Henry1931, Huckel1924, Wiersema1966, OBrien1978}, and they were experimentally verified \cite{Vorwerg1997}.
The theory of surface conduction \cite{Bikerman1933, Saville1983} also enables the extraction of the surface charge density from conductivity measurements \cite{Stein2004, Bocquet2016}.

It is not clear when ion-dissociable groups and ion adsorption at the interface were recognized as the main causes of surface electrification.
The quantitative adsorption isotherm of ions was derived using statistical mechanical theory at the beginning of the 20th century \cite{Langmuir1918, Stern1924, Grahame1947}.
These theories were applied to characterize the surface charge measured by potentiometric titration \cite{Borkovec2001}.
In the middle of the 20th century, Soviet and Dutch groups independently developed the theory of colloid stability in solutions, called the Derjaguin-Landau-Verwey-Overbeek (DLVO) theory \cite{DerjaguinLandau1941,VerweyOverbeek1948}.
The DLVO theory characterizes the interaction between the charged surfaces by van der Waals attraction and double-layer repulsion. 
This interaction can be measured by the disjoining pressure measurement of liquid films \cite{Derjaguin1939,Klitzing2009,Schelero2011}, surface force apparatus \cite{Israelachvili2010}, and atomic force microscopy with a colloidal probe \cite{Ducker1991, Butt1991}.
The double-layer force measurements can deduce the surface charge density of the water interface, as well as potentiometric titration.

Although the early history of the charged water interface starts from solid/water interfaces, the interfaces of gas/water and oil/water are also important systems for analytical and environmental chemistry.  
For gas/water and oil/water interfaces, surface tension measurement is a unique method to detect physical ion adsorption \cite{OnsagerSamaras1934, JonesRay1935, Washburn1930, PegramRecord2007, Uematsu2018, Abramzon1993, Marcus2010, WeissenbornPugh1996}.
Because the Gibbs adsorption isotherm is a thermodynamic relation, the relative surface affinity extracted from the surface tension data is quite reliable.
The air/water and hydrocarbon/water interfaces are normally inert, and these interfaces are classified as hydrophobic surfaces.
One of the debated issues of aqueous interfaces is the origin of surface charge at hydrophobe/water interfaces \cite{TseVoth2015,Vacha2008,BaiHerzfeld2016,Duignan2015,Baer2014}.  
To detect the surface charge of air/water interfaces, novel surface-sensitive analytical methods have been developed, such as vibrational nonlinear spectroscopy \cite{Shen2006,Shen2008, Shen2016,Tahara2009,Richmond2004} and mass spectrometry \cite{Colussi2006,Enami2010}.
Sum-frequency generation (SFG) spectroscopy is a surface-sensitive nonlinear optical method to detect the structure of a water interface. 
This method measures the second-order polarizability induced by the molecular vibrations at the interface. 
For the air/electrolyte solution interface, the SFG spectrum of the O--H stretching region (2800--4000$\,$cm$^{-1}$) is affected by surface-active ions, which allows us to characterize the surface affinity of ions from the difference between the SFG spectra of pure water and electrolyte solutions. 
Recently, the effect of oriented waters in the diffuse layer on second-order polarizability has been understood by sophisticated theory and experiments \cite{Shen2016}.
This allows us to extract the surface charge density from the SFG spectra of the charged water interface \cite{Shen2016}. 
Electrospray ionization mass spectrometry (ESI-MS) is another powerful tool to identify the amount and molecular species at the interface.
Although it is still not clear why ESI-MS is surface-sensitive, the interest in ESI-MS for a surface-sensitive analytical method is growing \cite{Mishra2019,Zhang2021}. 

Various characterization methods applied to the same surface yield a quantitative mismatch of the surface charge densities by different methods, which is an important issue in interface science \cite{LyklemaOverbeek1961,Saville1983,Saville1985,BonthuisNetz2012,BonthuisNetz2013,UematsuNetzBonthuis2017}.
Careful analysis of the quantitatively different surface charge densities of the same surface provides a deep understanding of the structure of the interface \cite{LyklemaOverbeek1961,Saville1983,Saville1985,BonthuisNetz2012,BonthuisNetz2013,UematsuNetzBonthuis2017,UematsuNetzBonthuis2018}.
The structure of the diffuse layer is universal for any material surface and ions, and it is well described by the classical Gouy-Chapman theory.
On the contrary, the structure of the outermost layer of the water interface has strong molecular individuality, and the description is inevitably material-specific and ion-specific. 
This interfacial layer was first modeled by Stern by introducing an immobile insulating layer with a thickness of the radius of absorbing hydrated ions \cite{Stern1924}.
Stern's picture of the interfacial layer is still quite useful to interpret the quantitative mismatches of the charge densities of the same surface using different methods.

Another important problem in interface science is the origin of the surface charge at interfaces.
Surface electrification usually occurs spontaneously at the interface.
The mechanisms of surface electrification are strongly ion-specific and material-specific.
Sometimes we cannot find any law or rule on surface electrification.
However, careful classification according to the types of ions and materials provides a universal picture of surface electrification, for example, the Hofmeister series \cite{Kunz2004}, and significant regulation by pH \cite{Borkovec2001}.

This review aims to provide a comprehensive overview of the state-of-the-art understanding of charged water interfaces.
Specifically, we summarize the characterization methods of surface charges and the mechanisms of charging at water interfaces.  
We believe that continuous efforts to investigate charged water interfaces will help in the fundamental progress of physics and chemistry at interfaces.

\section{Characterization of the charge at water interface}

First, we summarize various experimental methods to measure the surface charge of the water interface.

\subsection{Electrokinetics}

\begin{figure}[t]
\includegraphics{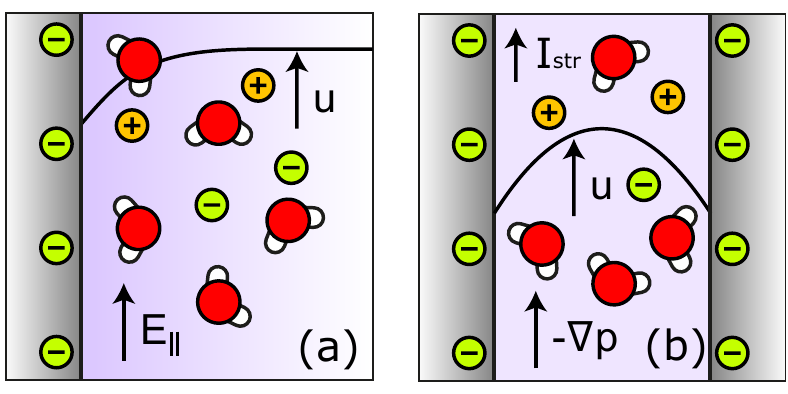}
\caption{Illustration of electro-osmosis and streaming current.
(a) Negatively charged solid surface with an electrolyte solution in semi-infinite space. 
The applied electric field $E_\parallel$ and the velocity profile $u$ are drawn. 
(b) Channel between two negatively charged solid surfaces is filled with an electrolyte solution.
The applied pressure gradient $-\nabla p$, parabolic velocity profile $u$, and streaming current $I_\mathrm{str}$ are drawn. 
}
\label{fig:4}
\end{figure}

Electrokinetics are the transport phenomena of liquids and charges driven by applied electric fields or pressure gradients.
Although many electrokinetic methods to characterize the surface charge are available, such as streaming potential, streaming current, and surface conductivity, 
the electro-osmotic/electrophoretic mobility measurement is the most widely used method to characterize the magnitude of the surface charge \cite{Hunter1981}. 
It is appreciable for any interface: solid/liquid, liquid/liquid, and gas/liquid interfaces.

First, we consider an ideal solid/liquid interface, where the solid surface is flat and charged homogeneously with the surface charge density $\sigma_0$. 
The liquid includes cations and anions whose valence is $q_i$ ($i=\pm$).
The solution has homogeneous viscosity $\eta$ and homogeneous relative dielectric constant $\varepsilon$.
When the external electric field $\boldsymbol{E}=E_\parallel \boldsymbol{e}_\parallel$ is applied in the direction parallel to the surface, as shown in Fig.~\ref{fig:4}a, the hydrodynamic equation for the steady state is given by
\begin{equation}
\frac{d}{dz}\left[\eta\frac{du}{dz}\right]=\rho(z)E_\parallel,
\label{eq:1}
\end{equation}  
where $u(z)$ is the velocity profiles of the solution.
The ionic charge density $\rho(z)$ satisfies the Poisson equation,
\begin{equation}
\frac{d}{dz}\left[\varepsilon\varepsilon_0\frac{d\psi}{dz}\right]=-\rho(z),
\label{eq:2}
\end{equation}
where $\varepsilon_0$ is the vacuum permittivity and $\psi(z)$ is the local electrostatic potential.
Combining eqs.~\ref{eq:1} and \ref{eq:2}, with no-slip boundary $u(0)=0$ at the interface $z=0$, the zero shear condition at bulk $du/dz|_{z\to\infty}=0$, constant potential condition $\psi(0)=\psi_0$, and zero potential at bulk $\psi(z)|_{z\to\infty}=0$, and the electro-osmotic mobility is calculated as
\begin{equation}
\frac{u(z)|_{z\to\infty}}{E_\parallel}= -\frac{\varepsilon\varepsilon_0\psi_0}{\eta}.
\label{eq:3}
\end{equation}
Eq.~\ref{eq:3} is called Helmholtz-Smoluchowski equation. 
To derive eq.~\ref{eq:3}, the assumption of the Boltzmann distribution for the ions is not necessary.
In experiments, the surface is not ideal.
The zeta potential of the surface is defined as 
\begin{equation}
\zeta = -\frac{\eta}{\varepsilon \varepsilon_0} \frac{u_\infty}{E_\parallel},
\label{eq:4}
\end{equation}
where $u_\infty$ here is the experimentally measured electro-osmotic velocity.
The zeta potential is considered to be the electrostatic potential at the plane where the no-slip boundary is imposed.
However, from a molecular viewpoint, it is not clear where the no-slip plane in a real surface is.
In addition, the Helmholtz-Smoluchowski equation neglects the effects of characteristic interfacial structures, such as the adsorption layer and packing of ions \cite{UematsuNetzBonthuis2018}, dielectric anomalies of interfacial water \cite{BonthuisNetz2013}, and slippage \cite{Sendner2009}.   
Therefore, the surface potential $\psi_0$ and the zeta potential $\zeta$ generally differ, which is one of the reasons the potential deduced from electro-osmosis is called zeta potential, not surface potential.  

For the electrophoresis of spherical charged solid particles, the electrophoretic mobility $\mu$ is no longer linear with respect to $\psi_0$, and the relation is given by \cite{Wiersema1966,OBrien1978},
\begin{equation}
\mu = f(\kappa R, \psi_0, \lambda_+, \lambda_-)\frac{\varepsilon\varepsilon_0\psi_0}{\eta},
\end{equation}
where $\kappa^{-1}$ is the Debye length, $R$ is the particle radius, $\lambda_\pm$ is the molar conductivity of cations and anions, and $f(\;)$ is a model-dependent function.
Analytical expressions of $f$ are known only for limiting cases. 
For solid particles, $f = 1$ for the planar limit $\kappa R\to \infty$, corresponds to Eq.~\ref{eq:3} while $f = 2/3$ for the point limit $\kappa R \to 0$, which is the H\"uckel formula.
For arbitrary $\kappa R$ and $\psi_0$, $f$ can be calculated numerically \cite{Wiersema1966,OBrien1978}.
Note that when $|\psi_0|\gtrsim 150\,$mV, the electrophoretic mobility can no longer be a monotonic function of $\psi_0$, and such a non-monotonic behavior of the electrophoretic mobility is observed for polystyrene latex solution in low salinity \cite{Vorwerg1997,Kobayashi2008}.
Because the inversion of $\psi_0$ from the electrophoretic mobility is not straightforward for non-monotonic cases, some studies use Eq.~\ref{eq:4} to convert $\zeta$ from the electrophoretic mobility of any geometry and surface potential.

To obtain the surface charge density from the zeta potential, we need to solve the Poisson-Boltzmann equation: 
\begin{equation}
\varepsilon\varepsilon_0\frac{d^2\psi}{dz^2} = 2ec_\mathrm{s}\sinh\left(\frac{e\psi}{2k_\mathrm{B}T}\right),
\label{eq:PB} 
\end{equation}
where $k_\mathrm{B}$ is the Boltzmann constant, $T$ is the temperature, $c_\mathrm{s}$ is the bulk salt concentration, and $e$ is the elementary charge.
We assume that the salt is monovalent $(q_i=\pm 1)$ and is fully dissociated in the solution.
By combining the  boundary conditions, $\psi(0)=\psi_0$ and $\psi(z)|_{z\to\infty}=0$, using Eq.~\ref{eq:PB}, the analytic solution for $\psi(z)$ is derived as \cite{Gouy1910,Chapman1913},
\begin{equation}
\psi(z)=\frac{2k_\mathrm{B}T}{e}\ln\frac{1+\mathrm{e}^{-\kappa z}\tanh(e\psi_0/4k_\mathrm{B}T)}{1-\mathrm{e}^{-\kappa z}\tanh(e\psi_0/4k_\mathrm{B}T)},
\label{eq:psi}
\end{equation} 
where $\kappa^2 = 2e^2c_\mathrm{s}/\varepsilon\varepsilon_0 k_\mathrm{B}T$.
Then, the relation between the surface potential and surface charge density is given by:
\begin{equation}
\sigma_0 = \varepsilon\varepsilon_0\left.\frac{d\psi}{dz}\right|_{z=0}=\sqrt{8\varepsilon\varepsilon_0 k_\mathrm{B}Tc_\mathrm{s}}\sinh\left(\frac{e\psi_0}{2k_\mathrm{B}T}\right).
\label{eq:6}
\end{equation} 
Using Eq.~\ref{eq:6}, we can obtain the surface charge density from the experimentally measured zeta potential.
However, the zeta potential is a well-known parameter to characterize the surface charge rather than the surface charge density itself, and the zeta potentials of various interfaces have been reported in the literature. 
Practically, electrophoretic light scattering is a widely used method  to measure the zeta potential of colloidal solutions because commercial apparatus is available, as well as direct microscopic observation of colloid electrophoresis \cite{Tottori2019}. 

An alternative method to evaluate the zeta potential is the streaming current measurement, which is defined by the ionic current driven by an applied pressure gradient, as illustrated in Fig.~\ref{fig:4}b,
\begin{equation}
I_\mathrm{str}=h\int^{L/2}_{-L/2}\rho(z)u(z)dz,
\end{equation}
where $h$ is the slit height, $L$ is the slit width, and $u(z)$ is the pressure-driven velocity satisfying $\eta\nabla^2u(z)-\nabla p=0$.
By integrating by part, the streaming current is derived by
\begin{equation}
\frac{I_\mathrm{str}}{Lh}\approx\left[\varepsilon\varepsilon_0\psi(z)\frac{du}{dz}\right]^{L/2}_{-L/2}=-\frac{\varepsilon\varepsilon_0\psi_0}{\eta} (-\nabla p),
\label{eq:str}
\end{equation}
where we assume $\kappa L \gg 1$. 
Eq.~\ref{eq:str} is very similar to Eq.~\ref{eq:3}, which is not coincidental. 
Using Onsager reciprocal relations, the linear coefficient of the streaming current density $I_\mathrm{str}/Lh$ with the applied pressure gradient $-\nabla p$ equals the electro-osmotic mobility.
The conversion equation of $\zeta$ from the streaming current for a flat surface is  
\begin{equation}
\zeta = -\frac{\eta}{\varepsilon \varepsilon_0} \frac{I_\mathrm{str}/Lh}{(-\nabla p)}.
\label{eq:5}
\end{equation}
To evaluate the zeta potential of a microchannel, the streaming current measurement can be conducted with reversible Ag/AgCl electrodes and DC voltage, and it is normally easier to measure the electro-osmotic mobility of the channel \cite{Alice2020}.  

\begin{figure}[t]
\includegraphics{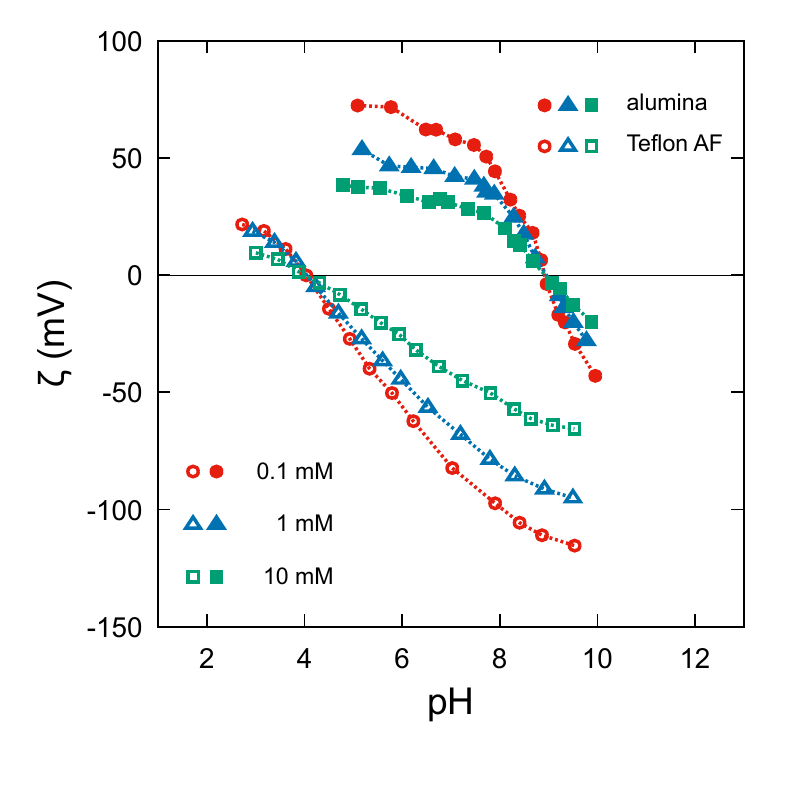}
\caption{Zeta potentials of alumina colloids \cite{Wiese1975_2} and flat Teflon AF surfaces \cite{Zimmermann2001} as a function of pH and salt concentration. Alumina data is obtained by electrophoresis measurement. The electrolyte was KNO$_3$, $\mathrm{pH}$ was adjusted by the addition of HNO$_3$ or KOH, and $T=25\,^\circ$C. For the conversion of $\zeta$ from the electrophoretic mobility, the Overbeek theory \cite{Wiersema1966} is used. The average radius was approximately $R=50\,$nm using an electron microscope. Teflon AF data is obtained by the streaming current measurement. 
The electrolyte was KCl, and the $\mathrm{pH}$ was adjusted by the addition of HCl or KOH. Because the surface is flat, $\zeta$ is directly converted from the streaming current via Eq.~\ref{eq:5}.
}
\label{fig:1}
\end{figure}

Fig.~\ref{fig:1} shows the zeta potential of titania and Teflon AF surfaces as a function of pH and salt concentration.
Both surfaces exhibit common behavior: $\mathrm{pH}$ of the isoelectric point ($\zeta=0$) is independent of the salt concentration, and at a pH lower than the isoelectric point, the zeta potential is positive, whereas at a pH higher than the isoelectric point, the zeta potential is negative. 

\begin{figure}[t]
\includegraphics{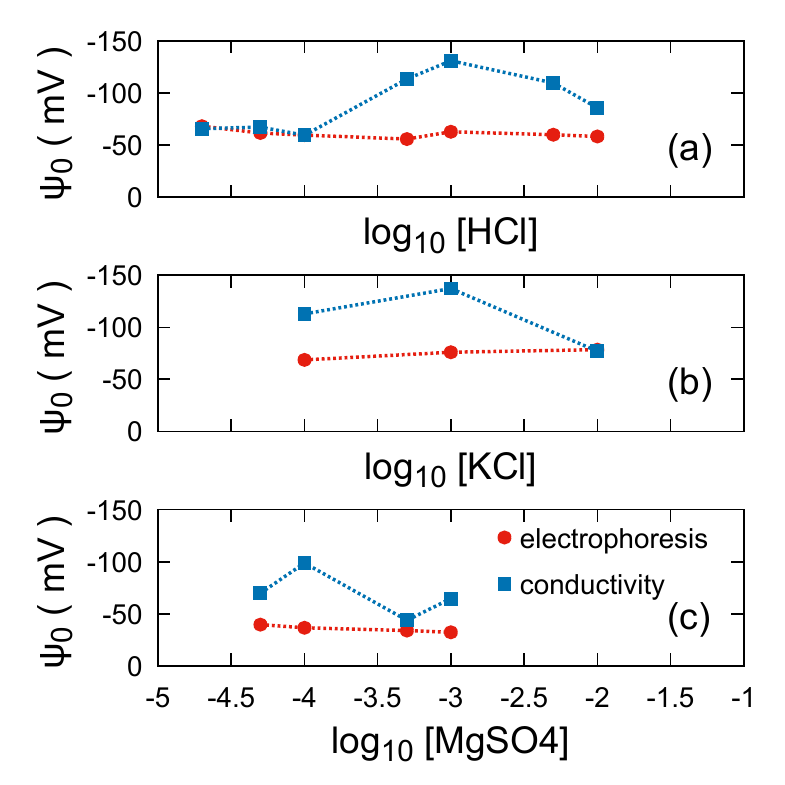}
\caption{Surface potentials of sulfonic polystyrene latex deduced from the electrophoretic mobility and surface conduction measurements \cite{Saville1985}. The radius was $83\,$nm, and $T=298$K was used. }
\label{fig:9}
\end{figure}

Another electrokinetic method to measure the surface charge is surface conduction measurements \cite{Bikerman1933, Saville1983, Saville1985, Stein2004}. 
Although the zeta potentials deduced from the electro-osmosis/phoresis and streaming current are identical, the surface potential deduced from surface conduction measurements is not necessarily identical to the zeta potential.
The surface conductivity $K_\sigma$ is the excess conductivity due to the surface charge, which is  defined by 
\begin{equation}
K_\sigma = \int^\infty_0 \sum_i\left[\lambda_i(c_i(z)-c_\mathrm{s})+eq_ic_i(z)\frac{u(z)}{E_\parallel}\right]dz,
\end{equation}
where $c_\pm$ is the local cation or anion density, $\lambda_i$ is the molar conductivity of the type $i$ ion, and $u(z)$ is the electro-osmotic velocity profile. 
To obtain a more explicit formula for the surface conductivity, we need to substitute the analytic solution of the Poisson-Boltzmann equation, Eq.~\ref{eq:psi}.
Here, we assume the equivalence of the molar conductivity of cations and anions ($\lambda=\lambda_+=\lambda_-$), and then, the explicit expression of the surface conductivity is
\begin{equation}
K_\sigma = \left(\frac{8\lambda c_\mathrm{s}}{\kappa}+\frac{16\varepsilon\varepsilon_0 k_\mathrm{B}T c_\mathrm{s}}{\kappa\eta}\right)\sinh^2\frac{e\psi_0}{4k_\mathrm{B}T}.
\end{equation}
This can be converted to the surface charge density using the relation $\sinh(2\mathrm{arsinh}x)=2x\sqrt{1+x^2}$ \cite{Bonthuis2012},
\begin{equation}
\sigma_0 = \frac{e\eta\sqrt{K_\sigma}}{\lambda \eta+2\varepsilon\varepsilon_0k_\mathrm{B}T}\sqrt{K_\sigma+\frac{8\lambda c_\mathrm{s}}{\kappa}+\frac{16\varepsilon\varepsilon_0 k_\mathrm{B}    T c_\mathrm{s}}{\kappa\eta}}.
\end{equation}
For a spherical solid surface, the conversion formula of the surface potential from the surface conductivity is more complicated \cite{Saville1983, Saville1985}. 
For planar and tube geometry, the surface conductivity measurement is useful to characterize the surface charge density \cite{Stein2004, SecchiNiguesJubinSiriaBocquet2016, Uematsu2018_JPC}.
Fig.~\ref{fig:9} shows surface potentials of sulfonic polystyrene latex colloid deduced from the electrophoresis and the surface conductivity measurements.
The surface potential of various electrolyte solutions reveals that the surface potential deduced from the surface conductivity is larger than the zeta potential deduced by electro-osmotic/phoretic or streaming current measurements \cite{Saville1985, Bonthuis2012}.
This has been explained by the counterion conduction in the interfacial layer, called anomalous surface conduction \cite{Dukhin1974}, and it was recently understood by zeta potential saturation \cite{Bonthuis2012}. 

\begin{figure}[t]
\includegraphics{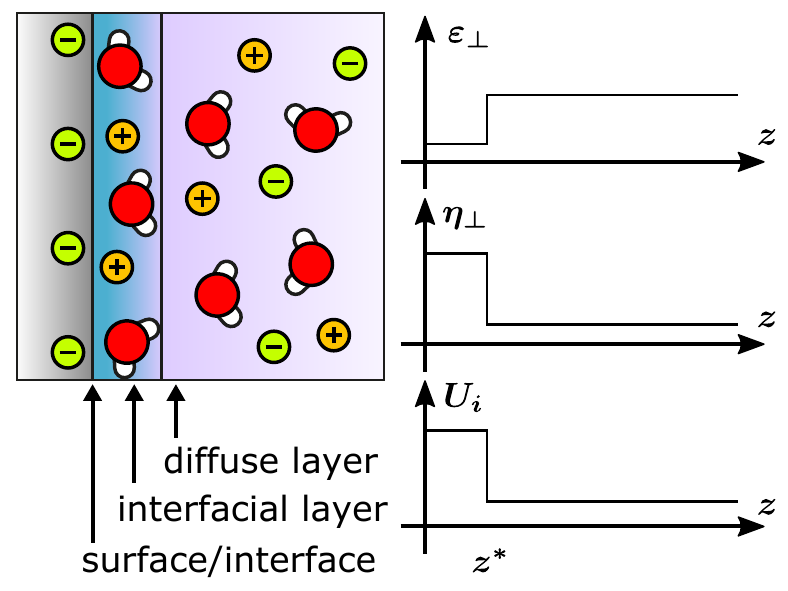}
\caption{
Schematic illustration of the analytical interfacial layer model \cite{UematsuNetzBonthuis2018}.
The left panel depicts an interfacial layer between a solid surface and a diffuse layer, where the local physical properties are significantly different from those in the bulk.
The right panel depicts the profiles of the dielectric constant, viscosity, and potential of the mean force along the normal coordinate $z$ used in the analytical model.
$z^*$ denotes the thickness of the interfacial layer.  
}
\label{fig:15}
\end{figure}

The mismatch of the surface potentials (or the surface charge densities) with different methods can be understood by introducing an interfacial layer into the model of the electric double layer \cite{LyklemaOverbeek1961,Saville1983,Saville1985,BonthuisNetz2012,BonthuisNetz2013,UematsuNetzBonthuis2017,UematsuNetzBonthuis2018}.
Fig.~\ref{fig:15} depicts the analytical interfacial layer model of the previous study where the thickness of the interfacial layer is set at a molecular dimension $z^*=0.5\,$nm \cite{UematsuNetzBonthuis2018}.
Even if the surface is not electrified, physical properties of the interfacial layer such as the dielectric constant $\varepsilon_\perp$ and viscosity $\eta_\perp$ are significantly different from those in the bulk phase. 
These interfacial anomalies significantly affect the differential capacitance, and hydrodynamic slip at the interface \cite{UematsuNetzBonthuis2018}.
Furthermore, the limited hydration ability of water in the interfacial layer induces ion adsorption at the interface.
This can be modeled by introducing the potential of the mean force into the Poisson-Boltzmann equation \cite{Horinek2007}.

\subsection{Potentiometric titration}

\begin{figure}[t]
\includegraphics{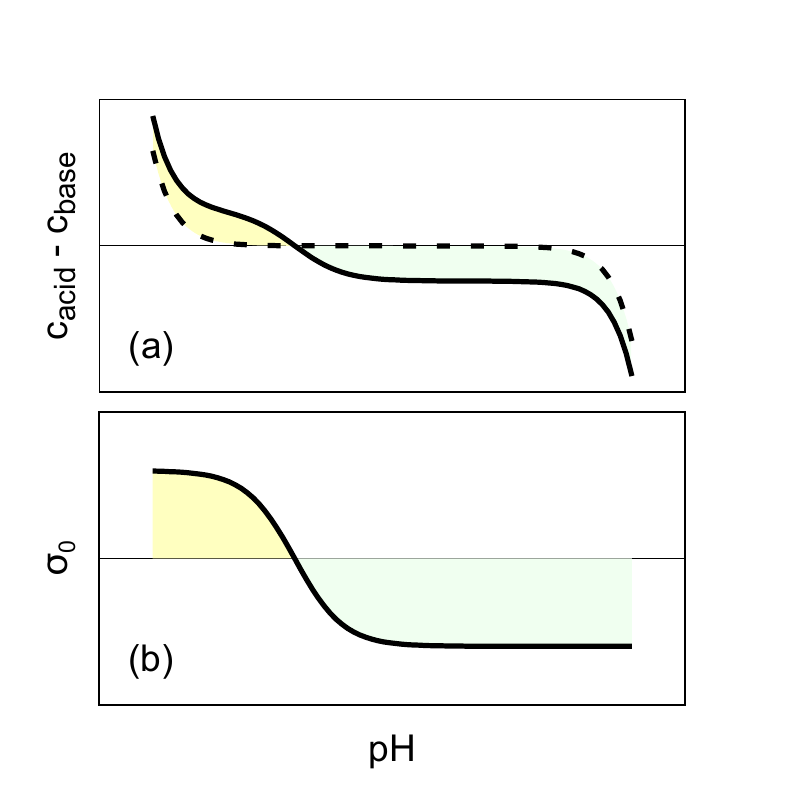}
\caption{Schematic plot of titration curves of strong acid with strong base \cite{Borkovec2001}.
(a). The titration curve of the blank acid solution (broken line) and titration curve of the acid solution with colloids (solid line).
(b). Surface charge density as a function of pH deduced from the difference between the two titration curves plotted in (a).}
\label{fig:5}
\end{figure}

Another method to characterize the surface charge density of colloids is potentiometric titration \cite{Borkovec2000, Borkovec2001, Tombacz2012, Lutzenkirchen2012}.
This method is based on stoichiometry, and thus, the dynamic properties of the surface do not affect the result.
In this sense, potentiometric titration is a more direct method than electrokinetic measurement.
However, potentiometric titration is limited to specific materials that satisfy the following two conditions: the electrification mechanism is already known, such as metal oxides \cite{Wiese1975_2}, carboxyl latex \cite{Borkovec2000, Borkovec2001}, and silver halides \cite{Kolthoff1936,Lyklema1978,Lyklema2007}, and a method for detecting the concentration of potential-determining ions is established, such as pH electrodes.

To explain the measurement principle of potentiometric titration, we used acid-base potentiometric titration as an example.
As a reference, the blank acid solution was first titrated by the strong base.
We need to record the initial pH and the concentration of acid. 
During titration, it is necessary to avoid the dissolution of carbon dioxide into the solution and to maintain a constant ionic strength by adding strong electrolytes.
A careful titration procedure gives the pH and acid cation concentration $c_\mathrm{acid}$, as a function of the added base anion concentration $c_\mathrm{base}$. 
The bulk charge neutrality condition yields
\begin{equation}
c_\mathrm{acid}-c_\mathrm{base}=c_\mathrm{H}(\mathrm{pH},I)-c_\mathrm{OH}(\mathrm{pH},I).
\label{eq:7}
\end{equation}
where $c^\mathrm{b}_\mathrm{H}$ and $c^\mathrm{b}_\mathrm{OH}$ are the bulk concentrations of $\mathrm{H^+}$ and $\mathrm{OH^-}$, respectively. 
The ionic strength $I$ is defined by
\begin{equation}
I=(1/2)\sum_iq_i^2c_i^\mathrm{b}
\end{equation}
where $c_i^\mathrm{b}$ is the bulk concentration of the type $i$ ions.
The added strong electrolytes did not change the difference between these two concentrations because they included the same amount of cations and anions. 
When the total ionic strength $I$ is small, these concentrations are ideally $c^\mathrm{b}_\mathrm{H}=10^{-\mathrm{pH}}\,$(M), and $c^\mathrm{b}_\mathrm{OH}=10^{\mathrm{pH}-14}\,$(M).
For high ionic strength, this relationship is not true, but these two concentrations are at least expected to depend only on $\mathrm{pH}$ and $I$. 
Eq.~\ref{eq:7} gives a titration curve as a function of $\mathrm{pH}$ (the broken line in Fig.~\ref{fig:5} (a)].

Then, a strong acid solution with colloids of interest was titrated using a strong base.
The charge neutrality condition leads to,
\begin{equation}
\tilde c_\mathrm{acid}-\tilde c_\mathrm{base}=c^\mathrm{b}_\mathrm{H}(\mathrm{pH},I)-c^\mathrm{b}_\mathrm{OH}(\mathrm{pH},I)+\sigma_0c_\mathrm{col}S
\label{eq:8}
\end{equation}
where $\sigma_0$ is the surface charge density determined by titration, $S$ is the specific surface area per colloidal mass, and $c_\mathrm{col}$ is the mass concentration of the colloids.
When we set the same $\mathrm{pH}$ and $I$, $c_\mathrm{H}$ and $c_\mathrm{OH}$ will be the same as those in Eq.~\ref{eq:7}.
However, $\tilde c_\mathrm{acid}$ and $\tilde c_\mathrm{base}$ differ from $c_\mathrm{acid}$ and $c_\mathrm{base}$ in Eq.~\ref{eq:7}. 
Therefore, the difference between the two titration curves, Eqs.~\ref{eq:7} and \ref{eq:8} yield the titration curves of the colloids,
\begin{equation}
\sigma_0 = \frac{1}{c_\mathrm{col}S}\left[\tilde c_\mathrm{acid}-c_\mathrm{acid}+\tilde c_\mathrm{base}-c_\mathrm{base}\right].
\label{eq:9}
\end{equation}

Fig.~\ref{fig:5} shows schematic plots of the potentiometric titration.
The broken line in Fig.~\ref{fig:5}a is the titration curve of the black acid solution, whereas the solid line represents the titration curves of the colloidal solution. 
The filled regions between the two curves shown in Fig.~\ref{fig:5}a are the excess charges due to the surface charge of the colloids.
Fig.~\ref{fig:5}b shows the surface charge density as a function of pH deduced from the titration, which is derived by the difference between the two titration curves plotted in Fig.~\ref{fig:5}a.

\subsection{Double-layer force measurement}
When two likely charged surfaces approach closely in an electrolyte solution, the diffuse double layers of each surface overlap, and the surfaces are repelled from each other because of electrostatic interactions.
The repelling force can be measured using various techniques, such as disjoining pressure measurements \cite{Derjaguin1992,Iyota2020,Schelero2011}, surface force apparatus \cite{Israelachvili2010}, atomic force microscopy with colloidal probes \cite{Ducker1991,Butt1991}, and optical tweezer trapping of two colloidal particles \cite{Grier1994}.
The first qualitative theory on the double-layer force was developed in the late 1930s \cite{Derjaguin1992, Langmuir1938-1}.
Since then, the disjoining pressure of surfactant-stabilized free-standing films (air/water/air) \cite{Yoon2009,Iyota2020}, and wetting films on a solid substrate (air/water/solid) \cite{Derjaguin1939,Read1969,Klitzing2009,Schelero2011} have been studied. 
A water film is stabilized by a balance of forces such as disjoining pressure, hydrostatic pressure, and capillary pressure \cite{Langmuir1938-1,Derjaguin1939,Read1969,Klitzing2009,Schelero2011}.
These films are classified according to their thickness into a Newton black film ($\alpha$-film) and a common black film ($\beta$-film).
The thickness of the first one is around a few nanometers stabilized by surfactants, whereas the thickness of the second one is around several tens to hundreds of nanometers, stabilized by a double-layer force. 
This section focuses on the common black films and the electrostatic contributions of the disjoining pressure.

The disjoining pressure $\Pi$ owing to the double-layer overlap was described by \cite{Behrens1999,Adar2018} 
\begin{equation}
\Pi(D)=2k_\mathrm{B}Tc_\mathrm{s}\left(\cosh\left(\frac{e\psi}{k_\mathrm{B}T}\right)-1\right)-\frac{\varepsilon\varepsilon_0}{2}\left(\frac{d\psi}{dz}\right)^2,
\label{eq:18}
\end{equation}
where $\psi(z)$ is the solution of the Poisson-Boltzmann equation between the likely charged surfaces separated by a finite thickness $D$.
For $D\gg\kappa^{-1}$, the disjoining pressure can be approximated by \cite{IsraelachviliBook}
\begin{equation}
\Pi(D)=64c_\mathrm{s}k_\mathrm{B}T\tanh^2\left(\frac{e\psi_0}{4k_\mathrm{B}T}\right)\mathrm{e}^{-\kappa D}.
\label{eq:dis}
\end{equation}
Combining eq.~\ref{eq:dis} with the measured disjoining pressure and film thickness $D$, the absolute value of the surface potential can be obtained.  

\begin{figure}[t]
\includegraphics{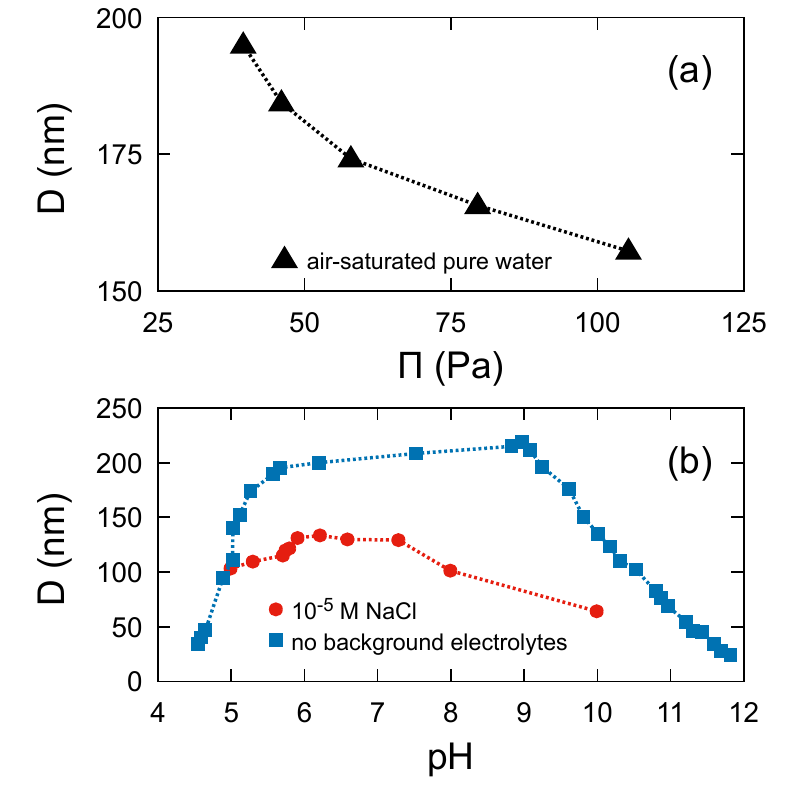}
\caption{
(a) Equilibrium thickness of air-saturated pure water as a function of the disjoining pressure $\Pi$ obtained using different ring radii \cite{Iyota2020}. 
(b) Equilibrium thickness of surfactant-free water film as a function of pH \cite{Yoon2009,Iyota2020}. 
The red circles are the data of $10^{-5}\,$M NaCl solution whose pH is adjusted by the addition of HCl or NaOH \cite{Yoon2009}.
The ring radius of a Scheludko cell is $R=2.0\,$mm; thus, the disjoining pressure is balanced with a capillary pressure of approximately $2\gamma/R =72.8\,$Pa.
The blue squares are the data of air-saturated pure water without background electrolytes whose pH is adjusted with the addition of HCl or NaOH \cite{Iyota2020}. 
The capillary pressure is $39.8\,$Pa with the ring diameter $2R=7.25\,$mm.
The lines are a guide for the eyes.
}
\label{fig:10}
\end{figure}

Very recently, surfactant-free air/water interfaces have been of great interest \cite{Klitzing2009,Yoon2009,Schelero2011,Iyota2020}.
Fig.~\ref{fig:10}a shows the thickness of a surfactant-free pure water film as a function of the disjoining pressure \cite{Iyota2020}, suggesting that the air/water interface is charged.
Fig.~\ref{fig:10}b shows the pH dependence of the film thickness with a constant disjoining pressure \cite{Iyota2020,Yoon2009}.
The two lines exhibit a similar trend, and the isoelectric point appears to be located at approximately $\mathrm{pH}=4$, which is in agreement with the zeta potential data of gas bubbles \cite{Yang2001}, and Teflon AF surfaces \cite{Zimmermann2001}.
Although these measurements do not reveal the sign of the interfacial charge, the thickness measurement for wetting films on the silica surface implies that the surfactant-free water interface is negatively charged \cite{Klitzing2009,Schelero2011}.

To measure the double-layer force at the solid/water/solid interface, a surface force apparatus and an atomic force microscope with colloidal probes have been developed \cite{Ducker1991, Butt1991,Israelachvili2010,Trefalt2020}.
Derjaguin approximation is very useful to connect the force measurement between curved surfaces to the disjoining pressure of flat surfaces \cite{IsraelachviliBook}.
For the two spherical colloid particles, where the separation is much smaller than the radii, the Derjaguin approximation yields 
\begin{equation}
\frac{F(D)}{R_\mathrm{eff}} = 2\pi W(D),
\end{equation}
where $F(D)$ is the observed force, $R_\mathrm{eff}^{-1}=1/R_1+1/R_2$ is the mean radius, and $W(D)$ is the interaction energy per unit area between two planar surfaces separated by $D$.
The interaction energy $W(D)$ is related to the disjoining pressure by an equation, 
\begin{equation}
W(D) = -\int^D_\infty \Pi(z)dz.
\end{equation}
Therefore, the measurement of the double-layer force $F(D)$ can directly relate to the disjoining pressure, Eq.~\ref{eq:dis} and the surface potential.

\subsection{Surface tension}

Surface tension measurement is not a direct method to determine the surface charge density, but it can detect the surface adsorption of ions.
In particular, this method is used for gas/liquid and liquid/liquid interfaces.
The surface tension increments $d\gamma$ can relate to the chemical potential increment $d\mu_i$ of type $i$ molecules by the Gibbs adsorption isotherm: 
\begin{equation}
d\gamma = -\sum_i \Gamma_i d\mu_i,
\label{eq:Gibbs}
\end{equation}
where the differential coefficient $\Gamma_i$ is the surface excess of the type $i$ molecules.
From eq.~\ref{eq:Gibbs}, we can conclude that the solutes that increase the surface tension (such as alkali halide salts) are repelled from the interface because the chemical potential $\mu_i(c_i)$ is usually an increasing function with respect to the bulk concentration $c_i$. 
In contrast, solutes that decrease the surface tension (such as surfactant) are attracted to the interface.
The concentration dependence of the ionic chemical potentials is not simple, and the activity coefficient of ions depends on both the ion type and ionic strength.
Nevertheless, the ideal gas approximation ($d\mu_i = k_\mathrm{B}Tdc_i^\mathrm{b}/c_i^\mathrm{b}$) is quantitatively reasonable \cite{Alexandre2020}.
 
\begin{figure}[t]
\includegraphics{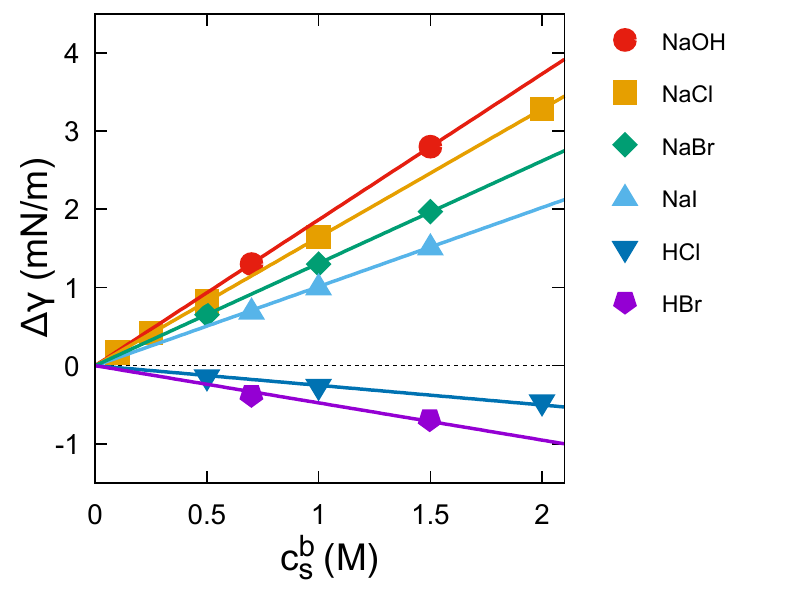}
\caption{Surface tension of the solutions of sodium salts and acids.
Original data are given in Ref. \citenum{Washburn1930}
The lines are linear fits.
}
\label{fig:2}
\end{figure}

Taking the surface tension of the air/water interface as an example, Fig.~\ref{fig:2} shows the surface tensions of the solutions of sodium salts and acids.
All the sodium salts shown in Fig.~\ref{fig:2} increase the water surface tension, indicating that these electrolytes are repelled from the air/water interface.
On the other hand, the acid solutions plotted in Fig.~\ref{fig:2} decrease the surface tension, meaning that the ions in the acid solution are attracted to the interface on average.   
Comparing the linear coefficients of the surface tension increase of the sodium salts, the anion surface affinities have an order of 
\begin{equation}
\textrm{(absorbing) }\mathrm{I^-}>\mathrm{Br^-}>\mathrm{Cl^-}>\mathrm{OH^-}\textrm{ (repelled)}.
\end{equation}
This order is very similar to the Hofmeister series for precipitation of protein solutions \cite{Kunz2004}, and it has been explained by the dehydration energies or hydrated ionic radii of ions \cite{Horinek2009}. 
For cations, the surface affinities have an order 
\begin{equation}
\textrm{(absorbing) }\mathrm{H_3O^+}>\mathrm{Na^+}\textrm{ (repelled)}.
\end{equation}
The surface tension analysis of various electrolyte solutions provides the relative surface affinity of individual ions because the surface tension difference is usually linear with the ion concentration. 

Although many models of ion partitioning to the interface have been suggested \cite{PegramRecord2007,Uematsu2018}, we used an analytical interfacial-layer model based on the Poisson-Boltzmann equation \cite{Uematsu2018}. 
An interfacial layer with the thickness $z^*=0.5\,$nm is introduced in the model.
In the interfacial layer, while the dielectric constant is modeled to be the same as the bulk one, ions are absorbed or repelled from the interface with the surface affinity $\alpha_i$ in units of $k_\mathrm{B}T$ for type $i$ ions.
The Poisson-Boltzmann equation of the model is given by
\begin{equation}
\varepsilon\varepsilon_0\frac{d^2\psi}{dz^2}=-e\sum_iq_ic_i^\mathrm{b}\mathrm{e}^{-eq_i\psi(z)/k_\mathrm{B}T-\alpha_i\theta(z^*-z)},
\label{eq:sf}
\end{equation}
where $\theta(z)$ is the Heaviside function. 
The boundary conditions in Eq.~\ref{eq:sf} are $d\psi/dz|_{z=0}=0$ and $\psi(z)|_{z\to\infty}=0$.   
The advantage of the model is that it is analytically tractable and takes many ion types into account in the same solution.
The surface excess of ions in the model is defined by
\begin{equation}
\Gamma_i = c_i^\mathrm{b}\int^\infty_0 \left[\mathrm{e}^{-eq_i\psi(z)/k_\mathrm{B}T-\alpha_i\theta(z^*-z)}-1\right]dz.
\end{equation}
From the potential and ionic profiles, the surface tension is calculated using the integral of Eq.~\ref{eq:Gibbs} or directly by the surface excess of the grand potential given by \cite{Uematsu2018} 
\begin{equation}
\Delta \gamma = -k_\mathrm{B}T\sum_i\Gamma_i -\int^\infty_0\frac{\varepsilon\varepsilon_0}{2}\left(\frac{d\psi}{dz}\right)^2 dz
\end{equation}
where $\Delta\gamma$ is the difference in surface tension from the pure water surface.
When the surface affinities of all the ions dissolved in the solution are quantitatively obtained, the surface charge density at the interface at $z=z^*$ is 
\begin{equation}
\sigma_0 = \int^{z^*}_0 \rho(z)dz \approx ez^*\sum_i q_ic_i^\mathrm{b}\mathrm{e}^{-eq_i\psi_0/k_\mathrm{B}T-\alpha_i}
\label{eq:adsorption}
\end{equation}
Combining Eq.~\ref{eq:6} using Eq.~\ref{eq:adsorption} gives the surface charge density at the air/water interface.

\subsection{Surface-sensitive nonlinear spectroscopy}

Surface-sensitive nonlinear spectroscopy is a unique tool to investigate the molecular structure of water interfaces \cite{Shultz2002,Shen2006}.
Second-harmonic generation (SHG) and SFG spectroscopy have a great advantage to detect interfacial structures because these spectroscopic methods measure the second-order polarizability $\chi^{(2)}$.
Because $\chi^{(2)}$ depends on the molecular orientation, the $\chi^{(2)}$ spectrum vanishes in the isotropic bulk phase.
At the interface, the isotropic symmetry is normally broken owing to the molecular orientation, and thus, the $\chi^{(2)}$ spectrum reflects the infrared (IR) vibrational spectrum of the molecules located at the interface.

\begin{figure}[t]
\includegraphics{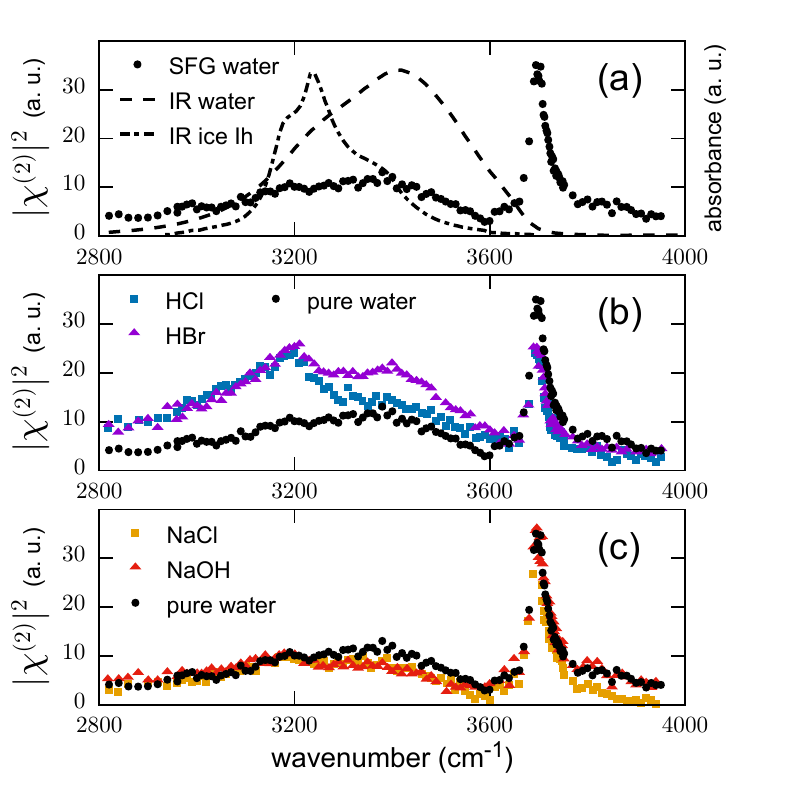}
\caption{
The IR and SFG spectrum of water, electrolyte, acid, and base.
(a). SFG spectrum of pure water (black circles) \cite{Mucha2005}, IR spectrum of bulk water (broken line) \cite{Bertie1996}, and IR spectrum of ice Ih phase (dashed-dotted line) \cite{Whalley1977}.
(b). SFG spectra of HCl (blue squares), HBr (purple triangles), and water (black circles) \cite{Mucha2005}.
(c). SFG spectra of NaCl (orange squares), NaOH (red triangles), and water (black circles)\cite{Mucha2005}.
The SFG spectra use the left axis (arbitrary unit), whereas the IR spectra use the right axis (arbitrary unit).
All the concentration of the electrolyte is $1.2\,$M. 
}
\label{fig:7}
\end{figure}

Fig.~\ref{fig:7} is the SFG spectra of pure water, acid, base, and electrolyte solutions in O--H stretching region ($2800$--$4000\,$cm$^{-1}$) \cite{Mucha2005}.
The SFG spectrum (ssp polarized) at the air/water interface shown in Fig.~\ref{fig:7}a by black circles exhibits a sharp band at $3700\,$cm$^{-1}$ and a broad band from $3000$ to $3600\,$cm$^{-1}$.
The sharp band ($3700\,$cm$^{-1}$) corresponds to free OH vibration, which has an upward OH orientation toward the air phase, which is unique to the SFG spectrum because the bulk IR absorbance spectrum (broken line) does not show any band in this range. 
The broad band ($3000$--$3600\,$cm$^{-1}$) is a hydrogen-bonded OH vibration, and two blunt peaks are located at $3200$ and $3400\,$cm$^{-1}$, respectively.
Because the small band at $3200\,$cm$^{-1}$ is observed in the IR spectrum of the bulk ice Ih phase (dashed-dotted line), it is called an ice-like structure.
On the other hand, the other small band at $3400\,$cm$^{-1}$ is called a liquid-like structure because the IR spectrum of bulk liquid water (broken line) exhibits a similar band at the same wavenumber.

The variations in the interfacial structure due to ion accumulation can be detected by the difference in the SFG spectra of the OH stretching region between the salt solution and pure water.
As molecular spectroscopy is not directly related to interface thermodynamics, the interpretation of the spectrum is not straightforward.
Fig.~\ref{fig:7}b shows the SFG spectrum of acid solutions: $1.2\,$M HCl (blue squares) and $1.2\,$M HBr (purple triangles) \cite{Mucha2005}.
These acids are known to be surface-active from the surface tension measurement.
Both spectra exhibit a decrease in the free OH band ($3700\,$cm$^{-1}$) and growth of ice-like ($3200\,$cm$^{-1}$) and liquid-like ($3400\,$cm$^{-1}$) bands compared to the pure water spectrum (black circles).
Because the SFG spectrum of the NaCl solution (orange points in Fig.~\ref{fig:7}c) does not exhibit significant difference from the pure water spectrum, the SFG spectra of the acids suggest that $\mathrm{H_3O^+}$ exists in the interfacial layer of acid solutions, and it perturbs the structure of OH stretching vibrations \cite{Mucha2005}. 
On the other hand, the spectrum of the NaOH solution (red triangles) is shown in Fig.~\ref{fig:7}c is very similar to the spectra of pure water and NaCl solutions (orange squares), indicating that $\mathrm{OH^-}$ is not surface-active at the air/water interface \cite{Mucha2005}.

\begin{figure}[t]
\center
\includegraphics{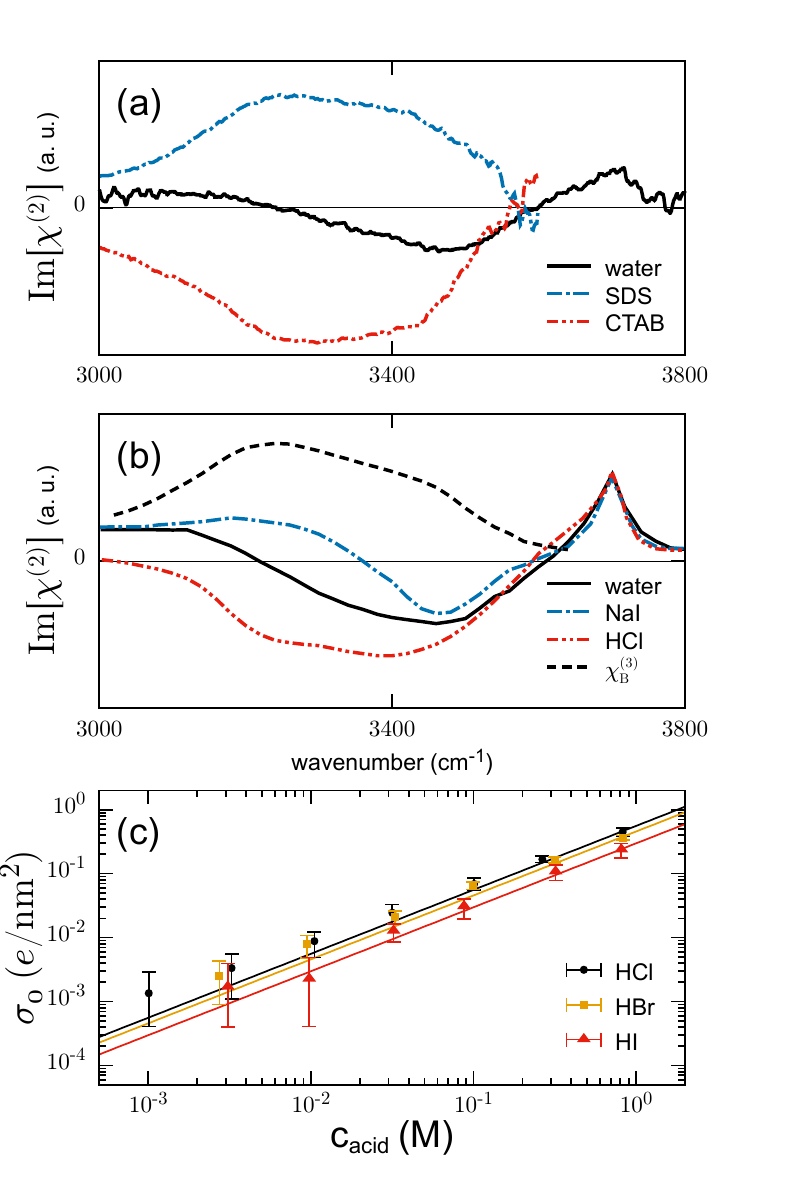}
\caption{
$\mathrm{Im}[\chi^{(2)}]$ of water, ionic surfactants, and electrolyte solutions obtained by PS-SFG spectroscopy.
(a). $\mathrm{Im}[\chi^{(2)}]$ of water (black solid line), $10\,$mM SDS (blue dashed-dotted line), and $10\,$mM CTAB solutions (red dashed-doubley-dotted line) \cite{Tahara2009}.
(b). $\mathrm{Im}[\chi^{(2)}]$ of water (black solid line), $1.2\,$M HCl solution (red dashed-doubly-dotted line), and $2.1\,$M NaI solution (blue dashed-dotted line) \cite{Shen2008}. $\chi^{(3)}_\mathrm{B}$ of the bulk water was also plotted together by the black broken line \cite{Shen2016}. 
(c) The surface charge density of halide acid solutions obtained by careful analysis of PS-SFG spectra \cite{Wen2020}. 
The symbols denote HCl (black circles), HBr (orange squares), and HI (red triangles).
}
\label{fig:8}
\end{figure}

Recently, a more informative SFG method, phase-sensitive sum-frequency generation (PS-SFG) spectroscopy, has been developed.
This can measure $\mathrm{Re}[\chi^{(2)}]$ and $\mathrm{Im}[\chi^{(2)}]$, respectively, not the absolute intensity $|\chi^{(2)}|^2$, and this enables us to distinguish the orientation of the water dipole moment in the diffuse layer.
Fig.~\ref{fig:8}a shows PS-SFG spectra of pure water (black solid line), a 10mM sodium dodecylsulfate (SDS) solution (blue dashed-dotted line) and a 10mM cetyltrimethylammonium bromide (CTAB) solution (red dashed-doubly-dotted line) \cite{Tahara2009}.
The positive broad band at $3000\,$ to $3600\,$cm$^{-1}$ region in $\mathrm{Im}[\chi^{(2)}]$ corresponds to the OH stretching vibration oriented the hydrogen upwards.
Because SDS is a negatively charged surfactant, the air/water interface is covered by anionic  surfactants.
In the diffuse layer, the dipoles of the water molecules are oriented parallel to the static electric fields.
For the CTAB solution, negative broad bands appear in $\mathrm{Im}[\chi^{(2)}]$, indicating the water orientation of the double layer due to the positive surface charge of CTAB.  
The large difference between the $\mathrm{Im}[\chi^{(2)}]$ spectra of SDS and CTAB solutions is not distinct in the absolute intensity $|\chi^{(2)}|^2$ \cite{Tahara2009}.
Fig.~\ref{fig:8}b shows SFG spectra of air/electrolyte interfaces \cite{Shen2008}. 
The free OH band ($3700\,$cm$^{-1}$) does not differ significantly for both $2.1\,$M NaI solution (blue dashed-dotted line) and $1.2\,$M HCl solution (red dashed-doubly dotted line) from pure water (solid black line), whereas the broad band in $3000$ to $3600\,$cm$^{-1}$ varies positively or negatively from the spectrum of pure water \cite{Shen2008}.
Because the positive deviation means the negative surface charge density,
the hydronium ion concentration dominated the chloride ions at the outermost interfacial layer of the water surface.
Conversely, $\mathrm{I^-}$ is more surface-active than $\mathrm{Na^+}$ from the PS-SFG spectrum of the NaI solution \cite{Shen2008}. 

To extract the surface charge density from the PS-SFG spectra, the bulk third-order polarizability $\chi^{(3)}_\mathrm{B}$ effect on the $\chi^{(2)}$ spectrum has been recently investigated \cite{Shen2016}.
In the theory, the second-order polarizability is modified as 
\begin{equation}
\chi^{(2)} = \chi^{(2)}_\mathrm{S}+\chi^{(3)}_\mathrm{B}\boldsymbol{e}_z\int^\infty_0  E_0(z')\mathrm{e}^{i\Delta k_zz'} dz', 
\end{equation}
where $\chi^{(2)}_\mathrm{S}$ is the interfacial second-order polarizability, $\chi^{(3)}_\mathrm{B}$ is the bulk third-order polarizability, $E_0(z)$ is the static electric field due to the surface charge, and $\Delta k_z$ is the phase mismatch depending on the geometry of the SFG experiments.
In Fig.~\ref{fig:8}b, the $\chi^{(3)}_\mathrm{B}$ spectrum of bulk water deduced from the PS-SFG experiment is plotted as a black broken line in an arbitrary unit together with the $\chi^{(2)}$ spectra of the electrolyte solutions.  
Combining the Gouy-Chapman theory described by Eq.~\ref{eq:6} and careful analysis of the PS-SFG spectra \cite{Shen2016}, the surface charge density of the water interface can be determined by experimental PS-SFG spectra \cite{Shen2016}.
Fig.~\ref{fig:8}c shows the surface charge density of halide acid solutions deduced by the PS-SFG spectroscopy measurements \cite{Wen2020}.
From the linear fitting of the adsorption isotherm, the adsorption energy of the proton was deduced to be $-3.8\,$kJ/mol$=-1.5 k_\mathrm{B}T$, which is quantitatively similar to that extracted from the surface tension measurements $-0.9k_\mathrm{B}T$ \cite{Wen2020}.
However, their analysis neglected the surface potential effect on the ion partition to the interface, which is the factor, $\mathrm{e}^{-eq_i\psi_0/k_\mathrm{B}T}$, in Eq.~\ref{eq:adsorption}.

\subsection{Surface-sensitive mass spectrometry}

Mass spectrometry is a powerful method to identify the ion types and the relative amount of ions in the solution. 
It provides a spectrum of the mass-to-charge ratio of ions.
One of the ionization methods for mass spectrometry is electrospray ionization, and it has been revealed that sampling of electrospray ionization is surface-sensitive \cite{Colussi2006, Enami2010, Schroder2012, Beauchamp2015}. 
Microdroplets produced by these methods have excess charges in themselves, and after evaporation of the solvent, the droplet radius reduces to the Rayleigh limit of the Coulomb explosion. 
Afterward, some of these ions become naked ions without any solvent molecules, and they are detected by mass spectrometry. 
The ion signals detected by mass spectrometry are considered to be correlated with the charge at the air/water interface rather than at the bulk because of the process of sub-microdroplet formation \cite{Colussi2006, Zilch2008}.
Studies on surface-sensitive mass spectrometry often use an electrically grounded spray nozzle to generate microdroplets \cite{Colussi2006,Enami2010}.
Electrospray ionization refers to naked-ion formation using a voltage-applied nozzle and an evaporation-induced Coulomb explosion.
In this sense, ionization with a grounded nozzle is, strictly speaking, not electrospray, but sonic spray \cite{Hirabayashi1994}.
Nevertheless, in this review, we refer to it as electrospray ionization because it uses an evaporation-induced Coulomb explosion.
In this section, we describe two important experiments related to the surface charge characterization of water interfaces \cite{Colussi2006, Enami2010}. 

\begin{figure}[t]
\center
\includegraphics{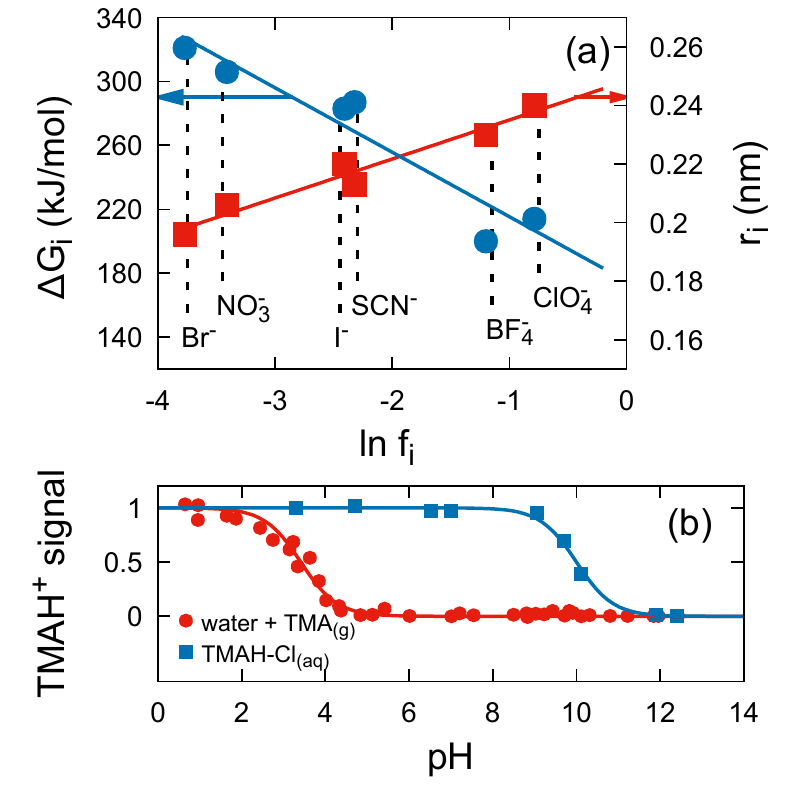}
\caption{
(a) Relative intensity of the anionic signal of electrospray ionization spectrometry $f_i$ \cite{Colussi2006} vs. anionic radii $r_i$ (red squares) \cite{MacusBook} and dehydration energies $\Delta G_i$ (blue circles) \cite{MacusBook}.
The relative intensities were converted from the signal intensity of each ion by electrospray ionization mass spectrometry using Eq.~\ref{eq:ms}.
The lines are linear fits of the data.
The solution was a 100 $\,\mu$M aqueous solution of the sodium salts NaSCN, NaNO$_3$, NaBr, NaBF$_4$, NaClO$_4$, and NaI. 
(b) Signal intensity of protonated trimethylamine ion (TMAH$^+$) deduced from electrospray ionization mass spectrometry as a function of pH at $1\,$atm and $293\,$K \cite{Enami2010}.  
The red circles are the intensities on water microjets exposed to 1.0-3.0 ppmv trimethylamine (TMA) gas, whereas the blue squares are the intensities on aqueous TMAH-Cl microjets in pure N$_2$ gas.
Both intensities are normalized to a unit signal intensity at $\mathrm{pH}=1$.
The lines are $(1+10^{\mathrm{pH}-\mathrm{pK_a}})^{-1}$, where $\mathrm{pK_a}$ are 3.4 and 10.0. 
}
\label{fig:11}
\end{figure}

Fig.~\ref{fig:11}a demonstrates the correlation between the relative anion signals of mass spectrometry (horizontal axis) and the dehydration energies (blue circles) as well as the ionic radii (red squares) \cite{Colussi2006}.
The test solution in Fig~\ref{fig:11}a is an equimolar solution of the sodium salts NaSCN, NaNO$_3$, NaBr, NaBF$_4$, NaClO$_4$, and NaI, where each concentration is 100$\,\mu$M.
The relative anion signal is defined by
\begin{equation}
f_i = \frac{I_i}{\sum_i I_i},
\label{eq:ms}
\end{equation}
where $I_i$ is the signal intensity of each anion.
Because the anion affinities to the air/water interface are correlated with the dehydration energy and ionic radii (see the section on surface tension), the results suggest that the sampling of the solution in ESI-MS is surface sensitive \cite{Colussi2006}.
Assuming that $I_i$ is proportional to the surface concentration of each ion, $c_i^\mathrm{b}\mathrm{e}^{-\alpha_i-eq_i\psi_0/k_\mathrm{B}T}$, the logarithm of the affinity $\ln f_i$ can be expressed as $\ln f_i = -\alpha_i +C$, where $C$ is independent of the anion type.  
Therefore, the negative correlation between $\ln f_i$ and $\Delta G_i$ is reasonable.
The positive correlation between $\ln f_i$ and $r_i$ indicates that larger ions require less dehydration energy, which can be explained by the Born theory of solvation. 
Fig.~\ref{fig:11}a proves that the relative intensity of anion in ESI-MS strongly correlates with the ionic radii and dehydration energy, i.e. the surface affinity, and this proves that ESI-MS is a useful method to detect the ion types accumulated at the surface. 

Fig.~\ref{fig:11}b shows the proton availability at the air/water interface as a function of bulk pH probed by protonated trimethylamine ion (TMAH$^+$) signal intensity of ESI-MS \cite{Enami2010}.
The blue squares are the signal intensities of the microjets of the TMAH-Cl solution with adjusted bulk pH using HCl and NaOH in N$_2$ gas.
The red circles represent the signal intensity of microjets of water with adjusted bulk pH in trimethylamine (TMA) gas.
The lines are the normal ionization curves, $(1+10^{\mathrm{pH}-\mathrm{pK_a}})^{-1}$, where $\mathrm{pK_a}$ are 3.4 (water microjets + TMA$_\mathrm{(g)}$) and 10.0 (TMAH-Cl$_\mathrm{(aq)}$ microjets). 
Experimental detail is described in Ref.~\citenum{Enami2010}. 
The signal of the TMAH-Cl$_\mathrm{(aq)}$ microjets seems to reflect the protonation equilibrium at the bulk, given by 
\begin{equation}
 \mathrm{TMAH}^+ \rightleftharpoons \mathrm{TMA_{(aq)}}+\mathrm{H^+_{(bulk)}},
\end{equation}
because the fitted $\mathrm{pK_a}=10.0$ is quite close to the literature value ($\mathrm{pK_a}=9.8$) of bulk TMAH$^+$ in water \cite{Enami2010}.
In contrast, the signal of water microjets with TMA$_\mathrm{(g)}$ gas exhibited a shifted $\mathrm{pK_a}=3.4$, which is significantly smaller than that of the TMAH-Cl$_\mathrm{(aq)}$ microjets. 
Assuming that the obtained TMAH$^+$ signal reflects the surface concentration of TMAH$^+$, the detected TMAH$^+$ ion seems to be formed by the protonation reaction at the interface, given by
\begin{equation}
\mathrm{TMA_{(g)}}+\mathrm{H^+_{(surface)}}\to \mathrm{TMAH}^+.
\end{equation}
This result demonstrates that the proton reactivity at the surface is weaker than that in bulk.
If the reactivity is just the concentration of the proton at the surface, the proton seems to be depleted from the interface under moderate and basic pH conditions. 
However, surface tension measurement, surface-sensitive nonlinear spectroscopy \cite{Saykally2005,Mucha2005}, and molecular dynamics simulations have demonstrated that the proton is surface-active at air/water interfaces \cite{Vacha2008,TseVoth2015,Duignan2015,Mamatkulov2017}.
Therefore, these studies qualitatively disagree with the ESI-MS experiment.  
To resolve these inconsistencies, many studies have been conducted so far \cite{Cook2002,Morita2013, Mishra2019, Enami2019, Mishra2019_2, Zhang2021}; however, this remains an open question.
The problem is the lack of quantitative theory to predict the signal intensity of ESI-MS in aqueous solutions.
The current understanding of microdroplet formation, water evaporation, and Coulomb explosion in ESI-MS is insufficient \cite{Fenn2000, Zlich2008, Konermann2010,Schroder2012, Beauchamp2015, Zhang2021}.   

\subsection{Other methods}

This section briefly mentions alternative experimental methods to characterize the surface charge density, namely, X-ray methods, electrochemical methods to measure the surface potential difference, and molecular dynamic simulations. 

X-ray is a powerful analytical tool to explore the electron structures of materials. 
X-ray photoelectron spectroscopy (XPS) can measure the surface potential by detecting the shift in the binding energy of the surface's specific electronic orbitals from the isoelectric point \cite{Bokhoven2013,Squires2016,Hemminger2017,Gladich2020,SoftXPS2018,Hemminger2019}.
X-ray absorption fine structure (XAFS) can detect surface-active heavy ions such as bromide ions and transition metal cations \cite{Watanabe1997,Takiue2003, Schlossman2015}.
X-ray reflectivity (XRR) can detect the profile of electron density along the interface \cite{Schlossmann2006, Schlossmann2013} .
Further explanation of these X-ray methods is beyond the scope of this review.

Measurement of the surface potential difference at the air/water interface is another unique tool to analyze surface charges \cite{Bonn2018, Nakahara2020, Allen2021}.
Compared to the zeta potential measurements, the surface potential difference measured by vibrating plate methods \cite{Bonn2018} or using radioactive electrodes \cite{Nakahara2020,Allen2021} includes the potential difference induced by the water orientation at the interface as well as ion adsorption.
It is still not clear how the surface potential difference is quantitatively related to the surface charge density measured by other methods, such as the zeta potential \cite{Bonn2018, Nakahara2020, Allen2021}.

Molecular dynamics simulations can directly study physical ion adsorption and the potential of the mean force of ions at the interface. 
However, classical molecular dynamics cannot describe chemical electrification involving a change in electronic chemical bonds, such as proton dissociation at carboxylic acids. 
In this sense, air/water and oil/water interfaces are quite easy to treat because the surface charge is mainly due to physical ion adsorption.
To obtain realistic ion adsorption behavior in molecular dynamics simulations, the choice of the force field of ions is essential \cite{Horinek2009,Fyta2012,Shavkat2013, Shavkat2016,Shavkat2018_2,Shavkat2018}.  
The force field parameters of ions installed in MD software or popular force field packages are usually not optimized for studying the interfacial properties of water interfaces.
Normally, a careful choice of these parameters using specific optimization procedures is necessary.
Polarizable force fields and ab-initio molecular dynamics are other choices. 
The former can reproduce the adsorption behavior of large halide ions at the air/water interface, and the latter can describe hydronium ions \cite{Mamatkulov2017}. 
However, ab-initio molecular dynamics simulations are still costly and limited to concentrated electrolyte solutions \cite{Chandra2014,Halonen2019}.
Based on the well-optimized force field, various physicochemical properties of the interface have been calculated by MD simulations, such as the SFG spectra of electrolyte solutions \cite{Morita2000, MoritaBook}, the surface tension of electrolyte solutions \cite{Mamatkulov2017}, and electrokinetic flow \cite{Joly2017}.
These results provide powerful insights to understand experiments from the molecular viewpoint.

\section{Electrification mechanisms}

\begin{table*}[t]
\begin{center}
\caption{Types of electrification and their examples.}
\label{tab:1}
\renewcommand{\arraystretch}{1.3}
\begin{tabular}{p{4.3cm}p{5.5cm}p{5.5cm}}
\\
{\bf Physical Ion Adsorption}\\\hline
material & origin & examples\\\hline
all materials \par (significant for inert surface) & ion adsorption from water phase \par owing to the hydration force & air, inert gas, hydrocarbon, \par perfluorocarbon\\\\
{\bf Chemical Electrification}\\\hline
material & origin & examples\\\hline
acidic and basic groups & proton dissociation and association & polymer with --COOH \cite{Borkovec2001}, --NH$_2$, --SO$_4^-$ \\ 
metal oxide/hydroxide & proton dissociation and association & SiO$_2$ \cite{Healy1992}, TiO$_2$ \cite{Wiese1975_2,Yates1980}, Al$_2$O$_3$ \cite{Wiese1975_2}, graphene oxide \cite{MLBocquet2020}\\
insoluble salt & ion dissociation and association & silver halides \cite{Preocanin2012,Preocanin2014}, CaF$_2$ \cite{Bonn2014,Bonn2020}\\
\end{tabular}
\renewcommand{\arraystretch}{1.0}
\end{center}
\end{table*}

In this section, we summarize the electrification mechanisms of interfaces. 
In Table~\ref{tab:1}, we tabulate types of electrification and their examples of materials. 
The mechanisms of electrification can be classified into two groups: physical ion adsorption and chemical electrification \cite{Parsons2019}.  
Physical ion adsorption means that the adsorbing ions are not bound to a specific site on the surface, and they do not drastically change the electronic structure of the surface or the ions, and thus, the chemical bonds are not newly created or broken.
The adsorption (or depletion) is caused mainly by the hydration force, which is the combined effect of van der Waals and electrostatic interactions between water and ions; thus, the classical (even nonpolarizable) MD simulation can describe physical ion adsorption \cite{Jungwirth2001}.
In contrast, chemical electrification is caused by the formation or breakage of chemical bonds, such as covalent and ionic bonds.
Because these reactions strongly depend on the electronic state of the surface and ions, chemical electrification is normally surface-specific and ion-specific. 
Physical ion adsorption occurs at all interfaces, but when a chemical electrification mechanism exists in a specific interfacial system, physical ion adsorption normally becomes minor compared to chemical electrification.
Therefore, physical ion adsorption is significant for chemically inert interfaces, such as air/water interfaces. 

\subsection{Physical ion adsorption}

Physical ion adsorption plays an important role in the electrification of interfaces.
In particular, because of the lack of a chemical electrification mechanism, chemically inert interfaces are electrified by ion adsorption.
Ion adsorption can be described by the potential of the mean force of ions as a function of the normal coordinate to the surface \cite{Horinek2007}.
When the cation and anion have the same potential, the surface is not electrified. 
However, when the surface affinities of the cations and anions differ, the surface is spontaneously electrified owing to the imbalance of the charge at the interfacial layer, regardless of whether the potential is attractive or repulsive. 
Therefore, physical ion adsorption (and depletion) causes surface electrification. 

The potential of the mean force for ions near the interface was first studied by Onsager in the 1930s \cite{OnsagerSamaras1934}.
They calculated the potential induced by the image effect due to the electrostatics of the interface between different dielectric media, given by \cite{OnsagerSamaras1934}
\begin{equation}
U_i(z) = \frac{\varepsilon-\varepsilon'}{\varepsilon+\varepsilon'}\frac{\mathrm{e}^{\kappa d_i}}{1+\kappa d_i}\frac{e^2}{\varepsilon\varepsilon_0 k_\mathrm{B}T}\frac{\mathrm{e}^{-\kappa z}}{4z},
\label{eq:OS}
\end{equation}
where $U_i(z)$ is the potential of the mean force in units of $k_\mathrm{B}T$, $\varepsilon$ is the dielectric constant of the solution, $\varepsilon'$ is the dielectric constant of the second phase, $\kappa^{-1}$ is the Debye length, $d_i$ is the ion diameter of type $i$ ion, and $z$ is the location of the ion from the interface at $z=0$. 
The force is independent of the charge sign of the ion, and it becomes stronger when the diameter is larger. 
For ions in water contacting high dielectric media such as metals, the potential provides an attractive force between the ions and the interface.
On the other hand, when the second phase is a low-dielectric medium such as air, the potential produces a repulsive force between the ions and the interface.
The surface tension of the electrolyte solution normally increases with concentration (see Fig.~\ref{fig:2}), and thus, the cations and anions are averagely repelled from the interface.
In fact, the experimental surface tension of the NaCl solution was quantitatively in good agreement with the theoretical prediction using Eq.~\ref{eq:OS} up to a concentration of $100\,$mM \cite{Uematsu2018,Alexandre2020}.
However, the ion-specific factor, $\mathrm{e}^{\kappa d}/(1+\kappa d)$ in Eq.~\ref{eq:OS}, is very weak, and it exhibits an opposite trend in the linear coefficient of the experimental surface tension of sodium halide solutions. 
Recent studies using molecular dynamics simulations have revealed that hydrophobic hydration i.e. the size of ions determines the interaction at the air/water interface \cite{Horinek2009}, and the image repulsion is not the main contribution to the potential of the mean force between the hydrophobic surface and ions \cite{Philip2018,Alexandre2020}.  

\begin{figure}[t]
\includegraphics{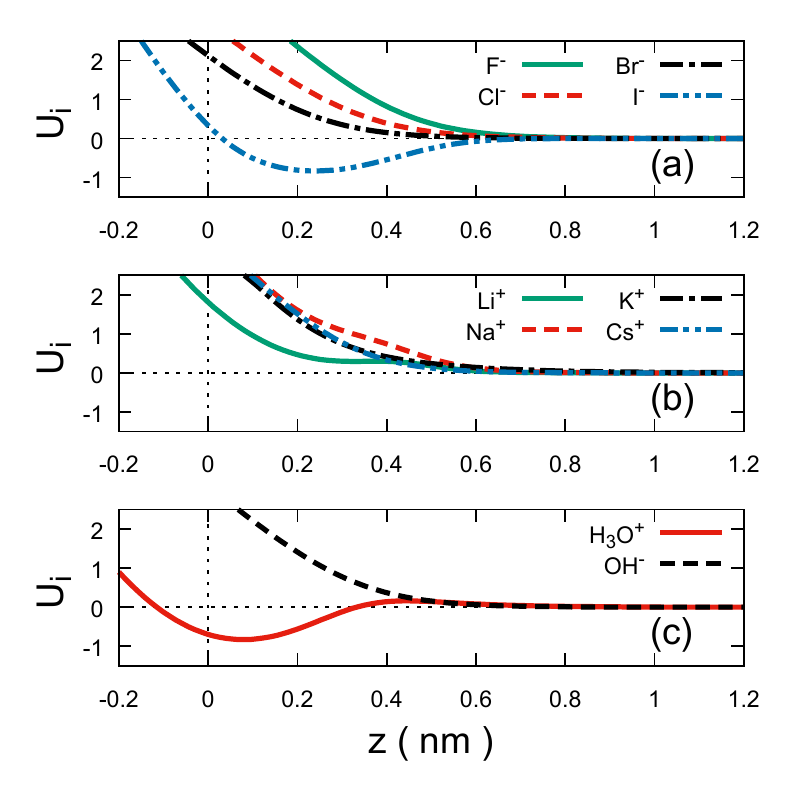}
\caption{
The potential of the mean forces of (a) halide ions \cite{Horinek2009}, (b) alkali metal ions \cite{Horinek2009}, and (c) water ions \cite{Mamatkulov2017} at the air/water interface.
The Gibbs dividing surface of water molecules is located at $z=0$.
The original data are fitted with analytical functions described in Appendix A.
The unit of the potential of the mean force is non-dimensionalized by $k_\mathrm{B}T$, and the temperature in the simulations was set to $T=300\,$K.
}
\label{fig:12}
\end{figure}

Hereafter, we focus on the surface affinity of ions at the air/water interface.
Fig.~\ref{fig:12} shows examples of the potentials of mean force of halide ions \cite{Horinek2009}, alkali metal ions \cite{Horinek2009}, and water ions \cite{Mamatkulov2017}.    
They are obtained by molecular dynamics simulation of SPC/E waters and ions with thermodynamically optimized nonpolarizable force fields \cite{Horinek2009,Horinek2009_JCP, Fyta2012,Shavkat2016}.
The potentials of the mean force derived by other ionic force fields are also available \cite{Dang2002,Mundy2009,Spoel2010}.
Because the potential of the mean force is related to the ionic concentration profile at the infinitesimal bulk concentration by $c_i(z)=c_i^\mathrm{b}\mathrm{e}^{-U_i(z)}$,
the surface excess of the type $i$ ions in the dilute limit is 
\begin{equation}
\frac{\Gamma_i}{c_i^\mathrm{b}}=\int^\infty_0 \left(\mathrm{e}^{-U_i(z)}-1\right)dz + \int^0_{-\infty} \mathrm{e}^{-U_i(z)} dz, 
\label{eq:PMF1}
\end{equation}
where the Gibbs dividing surface is located at $z=0$.
Using the analytical interfacial layer model \cite{Uematsu2018} and setting the interfacial layer thickness $z^*=0.5\,$nm, the surface affinity of each ion can be deduced by 
\begin{equation}
\alpha_i = -\ln\left(1+\frac{\Gamma_i/c_i^\mathrm{b}}{z^*}\right).
\label{eq:PMF2}
\end{equation} 
The ionic surface affinities to air/water interface deduced from the potentials of mean force plotted in Fig.~\ref{fig:12} are tabulated in the column of $\alpha_i^{\mathrm{(PMF)}}$ of Table \ref{tab:2}.

On the other hand, if the surface affinity of a reference ion ($\mathrm{Cl^-}$ in this study) is set from the reliable potential of the mean force in MD simulation, fitting a thermodynamic model with the experimental surface tension provides the surface affinity of the cation.
Using the analytical interfacial layer model \cite{Uematsu2018}, the cationic surface affinities to air/water interface $\alpha_i^\mathrm{(exp)}$ were determined as shown in Tab.~\ref{tab:2} \cite{Uematsu2018,Uematsu2019,UematsuCOLCOM, Uematsu2020}.  
In a similar manner, for other anions, the anionic surface affinities were obtained by a fit of the analytical interfacial layer model with fixing $\alpha_\mathrm{Na}=1.2$ \cite{Uematsu2018,Uematsu2019,UematsuCOLCOM, Uematsu2020}.
Comparing $\alpha_i^\mathrm{(PMF)}$ with $\alpha_i^\mathrm{(exp)}$, a strong qualitative correlation are observed.
This indicates that the well-optimized force fields qualitatively explain the ionic surface affinities deduced from the surface tension measurements.
In the following sections, we discuss the ion-specific adsorption of halide ions, alkali metal ions, water ions, and so on in more detail. 

\begin{table}[t]
\caption{The surface affinities of ions to the air/water interface.
$\alpha_i^\mathrm{(PMF)}$ were determined using the PMFs plotted in Fig.~\ref{fig:12} \cite{Horinek2009,Mamatkulov2017} using Eqs.~\ref{eq:PMF1} and \ref{eq:PMF2}, whereas $\alpha_i^\mathrm{(exp)}$ are determined by the fit of the analytical interfacial layer  model \cite{Uematsu2018} with the experimental surface tension of electrolytes \cite{Uematsu2018,Uematsu2019,UematsuCOLCOM, Uematsu2020}.
The surface affinities $\alpha_i^\mathrm{(exp)}$ for lithium and iodide ions are obtained in Appendix B.
}
\label{tab:2}
\begin{center}
\begin{tabular}{lrr}
\hline
Ion		&	$\alpha_i^{\mathrm{(exp)}}$&$\alpha_i^{\mathrm{(PMF)}}$\\\hline
$\mathrm{F^-}$	&	1.3		&1.90\\
$\mathrm{Cl^-}$	&	reference	&1.02\\
$\mathrm{Br^-}$	&	0.4		&0.55\\
$\mathrm{I^-}$	&	0.0		&$-0.66$\\
$\mathrm{Li^+}$	&	1.2		&0.52\\
$\mathrm{Na^+}$	&	1.2		&1.49\\
$\mathrm{K^+}$	&	1.2		&1.07\\
$\mathrm{Cs^+}$	&	1.2		&1.00\\
$\mathrm{H_3O^+}$&	$-0.9$		&$-0.68$\\
$\mathrm{OH^-}$	&	1.6		&1.01\\
$\mathrm{HCO_3^-}$&	$-0.4$		&---\\
$\mathrm{CO_3^{2-}}$	&1.4		&---\\
$\mathrm{C_{12}H_{25}SO_4^-}$&$-15.6$	&---\\ 
$\mathrm{C_{12}H_{25}(CH_3)_2NH^+}$&$-14.5$&---\\
$\mathrm{C_{12}H_{25}(CH_3)_3N^+}$&$-11.1$&---\\\hline
\end{tabular}
\end{center}
\end{table}

\subsubsection{Halide and alkali metal ions} 
\ \\
The difference between the surface affinities of halide ions was first suggested by surface tension measurements \cite{Washburn1930}.
The affinity order is known to be very similar to that of the Hofmeister effect on the salting-out of protein solutions \cite{Kunz2004}.
In the 2000s, molecular dynamics simulation studies using polarizable force fields revealed that chloride, bromide, and iodide ions were rich compared to sodium ions, whereas fluoride ions were repelled as sodium ions \cite{Jungwirth2001, Dang2002}.  
Subsequently, surface-sensitive nonlinear spectroscopy revealed an increase in the density of iodide ions, and some studies detected the enhancement of bromide ions at the air/water interface \cite{Liu2004,PetersenSaykally2004, Richmond2004}.
Although the quantitative estimate of the adsorption energies is still space to argue, the surface affinities of halide ions at the air/water interface exhibits the order
\begin{equation}
\mathrm{(absorbing)}\quad\mathrm{I^-}>\mathrm{Br^-}>\mathrm{Cl^-}>\mathrm{F^-}\quad\mathrm{(repelled)}.
\end{equation}
This trend agrees with the potentials of mean force plotted in Fig.~\ref{fig:12}a \cite{Horinek2009}.

Similar to halide ions, alkali metal ions exhibit repulsion from the air/water interface. 
However, the experimental surface tensions of chloride salts do not exhibit distinct differences in the repulsion strength, as shown in Table~\ref{tab:2} \cite{Washburn1930, PegramRecord2007, Uematsu2019}.    
This trend is also observed in the potentials of the mean force plotted in Fig.~\ref{fig:12}b \cite{Horinek2009}.
The potential of the mean force of $\mathrm{Li^+}$, however, shows anomalously weak repulsion \cite{Horinek2009,Hemminger2017}, even though lithium ions are the smallest ions among the alkali metal ions. 
This fact is related to the  large Stokes radius of lithium ions due to the strong bonding of hydration water molecules \cite{Horinek2009}.
 
So far, we have discussed ion adsorption at the air/water interface, i.e. the hydrophobic surface.
For charged or polar surfaces like protein solution and metal oxide surfaces, MD simulation revealed that the order of the ionic surface affinities is sometimes similar to reversed Hofmeister series \cite{Nadine2010, Nadine2013, Nadine2016}, which is in agreement with experiments of silica surface \cite{Sivan2016}.  

\subsubsection{Multivalent metal ions and other inorganic ions}
\ \\
Divalent ions such as calcium ($\mathrm{Ca^{2+}}$) and magnesium ($\mathrm{Mg^{2+}}$) are important in biology.
Thermodynamically optimized force fields for divalent ions are available \cite{Shavkat2013,Shavkat2018}, and the surface tensions of these salts have already been measured \cite{Washburn1930,WeissenbornPugh1996}.
However, their adsorption at the air/water interface is not fully understood \cite{MgCl2_2017}.  
The surface affinities of other divalent metal ions like Mn$^{2+}$, Fe$^{2+}$, and Cu$^{2+}$ were studied by surface tension measurement \cite{Washburn1930} and SFG spectroscopy \cite{Hemminger2019}.
Trivalent metal ions such as Fe$^{3+}$,  $\mathrm{Al}^{3+}$,  $\mathrm{La^{3+}}$, and $\mathrm{Y^{3+}}$, have been studied in the context of charge reversal and re-entrant phase transition in protein solutions \cite{Fajun2008, Lyklema2012, Fajun2017,  Fajun2018, Tahara2018, Kobayashi2019, Nd2019, Fajun2021, Allen2015}, but the surface affinity to the air/water interface remains unclear \cite{WeissenbornPugh1996}.
Some of these divalent and trivalent metal ions are transition metals, and they easily form ion complexes by coordinating various ligand molecules.
For example, the hydrolysis of cations (coordinating hydroxide ions as ligands) plays an important role in the charge reversal of colloids in solution \cite{Kobayashi2019, Lyklema2007}. 
The anisotropic and electronic interactions between multivalent metal ions and ligands have not been treated in classical molecular dynamics simulations because of the lack of reasonable force fields.

Oxyacid ions are another class of important ions in environmental and industrial chemistry.
NO$_x$ and SO$_x$ from the atmosphere dissolve in water, forming $\mathrm{NO_3}^-$ \cite{Jungwirth2003,Nitrate2020}, $\mathrm{HSO_4}^-$, and $\mathrm{SO_4}^{2-}$ \cite{Allen2015}. 
Phosphate ions ($\mathrm{PO_4}^{3-}$) cause eutrophication and hypochlorite ions (ClO$^-$) are used for the disinfection of drinking water. 
Ion adsorption and surface affinity of such oxyacid ions have been studied experimentally \cite{Hiemstra1999,Jungwirth2003,Sugimoto2018,Shavkat2018_2, Mondal2021}, whereas the force fields for oxyacid ions are still limited \cite{Shavkat2018_2}.

\subsubsection{Water ion}
\ \\
At bulk, water molecules spontaneously dissociate hydronium and hydroxide ions by the reaction:
\begin{equation}
2\mathrm{H_2O} \rightleftharpoons \mathrm{H_3O^+}+\mathrm{OH^-},
\end{equation}
with the ionic product $c_\mathrm{H_3O} c_\mathrm{OH}=\mathrm{K_w}=10^{-14}\,$(M$^2$). 
It has been debated for a long time which of  water ions are absorbed into air/water interfaces \cite{Roger2012,Vacha2012,Yan2018,Uematsu2019,Hassanali2020,Mishra2020,Barisic2021}. 

Many hydrophobic surfaces, such as gas bubbles, oil droplets, and fluoropolymer solid surfaces, exhibit negative zeta potential, and the absolute zeta potential increases as the pH increases. 
Thus, $\mathrm{OH}^-$ seems to be responsible for the negative zeta potential of the hydrophobic surfaces.
Quantitatively, the characteristic zeta potential $-50\,$mV gives the surface affinity of $\mathrm{OH^-}$ from $-18$ to $-20k_\mathrm{B}T$ \cite{Lutzenkirchen2008}, which is doubtful because the typical ionic surfactant SDS has a similar surface affinity \cite{Uematsu2018, Uematsu2020}.  

Based on the surface tension data of HCl and NaOH solutions, the surface affinity of protons is $-0.9k_\mathrm{B}T$ and that of $\mathrm{OH}^-$ is $1.6k_\mathrm{B}T$ (see Table~\ref{tab:2}).
Ab-initio or classical molecular dynamics simulation \cite{TseVoth2015, Mamatkulov2017} and continuous solvent model \cite{Duignan2015} support $\mathrm{H_3O^+}$ adsorption, and the surface affinity observed in the simulations agreed with the values based on the surface tension data (see Table~\ref{tab:2}).
Furthermore, surface-sensitive nonlinear spectroscopy detects evidence of $\mathrm{H_3O^+}$ adsorption \cite{Mucha2005,Saykally2005} and $\mathrm{OH^-}$ depletion \cite{Mucha2005}. 
Fig.~\ref{fig:12}c shows the potentials of mean force for $\mathrm{H_3O^+}$ and $\mathrm{OH^-}$ at the air/water interface derived a thermodynamically optimized force field \cite{Shavkat2016,Mamatkulov2017}.
The potential of $\mathrm{H_3O^+}$ exhibits a characteristic minimum at the interface, whereas the potential of $\mathrm{OH^-}$ is simply repulsive like the chloride ion plotted in Fig.~\ref{fig:12}a.

However, some studies revealed an opposite trend, suggesting that hydroxide ion is adsorptive to hydrophobic surfaces \cite{Beattie2004,Enami2010,Enami2012,Mbocquet2016,MBocquet2019}.
Surface sensitive mass spectrometry supports $\mathrm{OH}^-$ adsorption rather than $\mathrm{H_3O^+}$ adsorption at the gas/water interface \cite{Enami2010, Enami2012}. 
Potentiometric titration of surfactant-free hexadecane emulsion in water suggests that hydroxide ion is responsible for the negative surface charge of the hexadecane/water interface \cite{Beattie2004}.
Ab-initio molecular dynamics simulation has revealed the adsorption of hydroxide ions to graphene/water, and hexagonal boron nitride/water interfaces \cite{MBocquet2019}.

In summary, the surface affinity of water ions to hydrophobic surfaces has been intensely debated, and it is still not clear whether a universal mechanism governs the electrification of all hydrophobic materials.
Limiting the air/water interface, surface tension measurements and surface-sensitive nonlinear spectroscopy strongly support $\mathrm{H_3O^+}$ adsorption and OH$^-$ repulsion.

\subsubsection{Bicarbonate and carbonate ions}
\ \\
The dissolution of gas molecules plays an important role in atmospheric chemistry \cite{Jungwirth2004, Richmond2006}.
Inert gases such as $\mathrm{N_2}$, $\mathrm{O_2}$, and $\mathrm{Ar}$ do not react with water to form ions, but acidic gases such as carbon dioxide, nitrogen dioxide, and sulfur dioxide produce protons when they dissolve in water, causing acid rain.
This section summarizes the dissolution of carbon dioxide and the surface affinity of bicarbonate and carbonate ions at the air/water interface.

Dissolved carbon dioxide reacts with water by the following reactions:
\begin{equation}
\mathrm{CO_{2(aq)}}+\mathrm{H_2O} \rightleftharpoons \mathrm{H_2CO_{3(aq)}}
\end{equation}
\begin{equation}
\mathrm{H_2CO_{3(aq)}} + \mathrm{H_2O} \rightleftharpoons \mathrm{HCO_3^-} + \mathrm{H_3O^+}
\end{equation}
\begin{equation}
\mathrm{HCO_3^-} + \mathrm{H_2O} \rightleftharpoons \mathrm{CO_3^{2-}} + \mathrm{H_3O^+}
\end{equation}
Under ambient pressure, the concentration of bicarbonate ions in $\mathrm{CO_2}$-saturated water is approximately $2.5\,\mu$M, whereas the concentration of carbonate ions is $50\,$pM \cite{Persat2009-1}.
The surface affinity of carbonate and bicarbonate ions were studied by surface tension measurement \cite{Ozdemir2006}, sum-frequency generation spectroscopy \cite{Ozdemir2008,Allen2011}, and molecular dynamics simulations \cite{Ozdemir2008}.
Analyzing the experimental surface tension using the analytical interfacial layer model with $z^*=0.5\,$nm yields the surface affinities of $-0.4k_\mathrm{B}T$ for $\mathrm{HCO_3^-}$, and $1.4k_\mathrm{B}T$ for $\mathrm{CO_3^{2-}}$ (see Table~\ref{tab:2}) \cite{Uematsu2018,Uematsu2020}. 
This result indicates that $\mathrm{HCO_3^-}$ is weakly adsorptive and $\mathrm{CO_3^{2-}}$ is repulsive at the air/water interface, in agreement with SFG spectroscopy and molecular dynamics simulation \cite{Ozdemir2008, Allen2011}, whereas X-ray photoemission spectroscopy suggests an opposite trend, in which carbonate ions are absorbing stronger than bicarbonate ions \cite{Saykally2017}. 
 
Because the bulk concentration of bicarbonate ions is higher than the bulk concentration of the hydroxide ion under ambient conditions ($\mathrm{pH}=5.6$), it is, sometimes, considered to be the origin of the negative zeta potential of hydrophobic surfaces \cite{Med1991,Yan2018}.
Indeed, the bicarbonate ion is suggested to be surface-active \cite{Uematsu2020, Ozdemir2008, Allen2011}. 
However, the surface affinity of $\mathrm{HCO_3^-}$ is not sufficient to produce the magnitude of the experimental zeta potential \cite{Uematsu2020}.  

\subsubsection{Organic ions and ionic impurities}
\ \\
Ionic surfactants typically have both hydrophilic and hydrophobic groups, and their amphiphilic nature produces a large surface affinity to the air/water interface \cite{Tahara2009, Duignan2021}. 
Sodium dodecyl sulfate (SDS) is the most common anionic surfactant, and the critical micelle concentration in pure water is $8.2\,$mM \cite{Xu2013}. 
In the dilute limit, the solution exhibits a linear decrease in the surface tension.
The analytical interfacial layer  model with an interfacial layer thickness $z^*=0.5\,$nm gives the surface affinity of dodecylsulfate to the air/water interface as $-15.6 k_\mathrm{B}T$ (see Table~\ref{tab:2}) \cite{Uematsu2018}.
Such a strong surface affinity compared to small inorganic ions is caused by the hydrophobicity of the long alkyl chain.  

Carboxylic acids ($\mathrm{C_nH_{2n+1}COOH}$) are another common class of charged surface-active molecules in nature. 
A solution of carboxyl acids was studied using surface tension \cite{Washburn1930, Abramzon1993, Marcus2010, PegramRecord2007, Jungwirth2007, Allen2017}, MD simulation \cite{Jungwirth2007}, SFG spectroscopy \cite{Allen2017}, X-ray photoemission spectroscopy \cite{Werner2018}, and mass spectrometry \cite{Enami2016}.
These studies also reveal the shift of $\mathrm{pK_a}$ at the interface. 
The carboxyl acids exhibit $\mathrm{pK_a}=4.8$ in bulk in aqueous solutions \cite{Allen2017}, whereas at the air/water interface, $\mathrm{pK_a}$ increases as the carbon number increases ($n=7$, 8, 9) \cite{Allen2017}.     

Surface-active charged impurities are also a possible cause of surface electrification \cite{Roger2012,Uematsu2020,Karakashev2019}.
Needless to say, intentionally added ionic surfactants behave as an impurity \cite{UematsuCOLCOM}.
Here we discuss the importance of unspecified impurity in experiments \cite{Abe2019,Kulakov2002}. 
Normally, the absence of charged impurities in water is guaranteed by using ultrapure water produced by a combination of ion-exchange filtration and UV decomposition of organic impurities.
In addition, all glassware in the experiments should be cleaned using strong acid solutions such as piranha solution.    
However, all experimental studies are not performed with such an ideal condition.
Other contributions to the cleanliness of experiments, such as salt purity, oil purity, and gas purity, are sometimes not considered.
Therefore, obtaining and keeping a clean water interface is quite difficult.

A study on the zeta potential of hexadecane emulsions in water with different hexadecane purities revealed that the absolute zeta potential increased with increasing hexadecane purity \cite{Roger2012}.
The dependence of the zeta potential on the bulk pH is well explained by assuming proton-dissociable impurities with a single $\mathrm{pK_a}=5$ \cite{Roger2012}.
The fitted $\mathrm{pK_a}$ is close to one of the carboxyl acids, suggesting that the impurity in hexadecane is a fatty acid, and the zeta potential of the hexadecane/water interface is governed by the protonation/deprotonation of the impurity at the interface \cite{Roger2012, Roger2012_2, Uematsu2020}.
Recent studies on the zeta potential of emulsions with the accurate control of added impurity concentration and the level of cleaning process demonstrated that the zeta potential of surfactant-free emulsions could be reduced to $-10\,$mV \cite{Carpenter2019,Carpenter2020}. 
In addition, the characteristic minimum of the electrolyte surface tension, called the Jones-Ray effect, can be explained by assuming similarly charged impurities in water at nanomolar concentrations \cite{Uematsu2018, Duignan2018}.
The characteristic slip at the air/ultrapure water interface observed in the experiment \cite{Manor2008-2} suggests the presence of impurities at the ultrapure water interface. 

The impurity hypothesis is sometimes ruled out by various reasons \cite{Beattie2012,Roke2012,Roke2018,Roke2020}.
One of the reasons is that the behavior of zeta potential is quite reproducible and universal for various kinds of hydrophobic surfaces, and thus, why the same identity and concentration of impurities have occurred in different laboratories at different times \cite{Beattie2012}. 
However, careful literature examination reveals that the zeta potentials at the same ionic strength and the same material interface provided by different studies exhibit different magnitudes and isoelectric points \cite{Uematsu2020}, which supports the impurity hypothesis.

\subsection{Chemical electrification}

\begin{figure}[t]
\center
\includegraphics{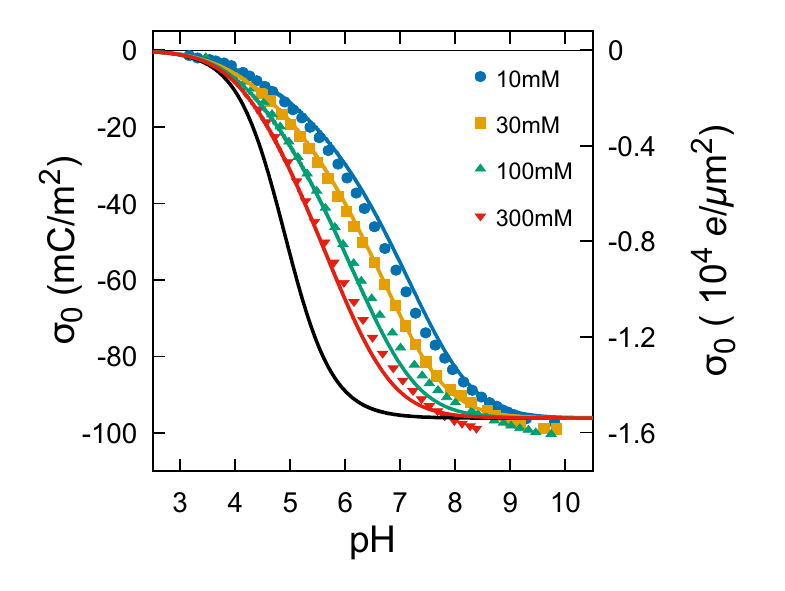}
\caption{Titrated surface charge density of carboxyl latex particles \cite{Borkovec2000}.
The colored lines represent the Langmuir adsorption isotherm, Eq.~\ref{eq:Langmuir}, with $\Gamma_0=0.6\,$nm$^{-2}$, and $\mathrm{pK_a}=4.9$. 
The black line represents the isotherm, eq.~\ref{eq:Langmuir}, with $\psi_0=0$.}
\label{fig:6}
\end{figure}

In this review, chemical electrification is defined as surface electrification, which changes the chemical bonds.
In Table~\ref{tab:1}, three types of materials governed by chemical electrification are tabulated. 
The electrification of these materials is described by very similar chemical equilibrium equations.  
The surface charge of most metal oxides and metal hydroxides in contact with water is governed by the association and dissociation of protons in the solution \cite{Borkovec2001,Trefalt2016}. 
The surface charge of polymeric materials and proteins is also governed by the charge regulation of acidic and basic groups on the surface.
For silanol groups on a silica surface, for example, the deprotonation reaction is given by
\begin{equation}
\mathrm{-SiOH} \rightleftharpoons \mathrm{-SiO^-} +\mathrm{H^+},
\end{equation}
and for carboxyl groups on a colloid surface, the reaction is 
\begin{equation}
\mathrm{-COOH} \rightleftharpoons \mathrm{-COO^-}+\mathrm{H^+}.
\end{equation}
For amine groups on a protein surface, the protonation reaction is 
\begin{equation}
\mathrm{-NH_2} + \mathrm{H^+} \rightleftharpoons \mathrm{-NH_3^+}.
\end{equation}
These deprotonation and protonation reactions at the electrification sites on the surface are described by the equilibrium equation for the chemical reaction.
In the following paragraph, the electrification of carboxylic latex is explained in detail.  

The acid dissociation constant $\mathrm{pK_a}$ of the carboxylic groups on the surface is defined as:
\begin{equation}
\mathrm{pK_a}=-\log_{10}\mathrm{K_a} = -\log_{10}\frac{\Gamma_\mathrm{COO}c_\mathrm{H}^\mathrm{int}}{\Gamma_\mathrm{COOH}}
\end{equation}
where $\Gamma_\mathrm{COO}$ and $\Gamma_\mathrm{COOH}$ are the surface densities of each group, and $c_\mathrm{H}^\mathrm{int}$ is the proton concentration near the interface. 
In the mean-field description, the proton concentration near the interface is given by $c_\mathrm{H}^\mathrm{int}=10^{-\mathrm{pH}}\mathrm{e}^{-e\psi_0/k_\mathrm{B}T}$, where $\psi_0$ is the surface potential and $k_\mathrm{B}T$ is the thermal energy. 
Using the surface density of the total carboxyl groups, $\Gamma_0=\Gamma_\mathrm{COO}+\Gamma_\mathrm{COOH}$, the Langmuir adsorption isotherm with the electrostatic effect was obtained.
\begin{equation}
\sigma_0 = -e\Gamma_0 \frac{\mathrm{K_a}10^{\mathrm{pH}}\mathrm{e}^{e\psi_0/k_\mathrm{B}T}}{1+\mathrm{K_a}10^{\mathrm{pH}}\mathrm{e}^{e\psi_0/k_\mathrm{B}T}},
\label{eq:Langmuir}
\end{equation}
where $\sigma_0$ is the surface charge density.  
Combining eq.~\ref{eq:Langmuir} using Eq.~\ref{eq:6}, the surface charge density and surface potential were obtained by the given $\mathrm{pH}$ and $c_\mathrm{s}$.
The evidence of this mechanism is that the fitting of the titration data of carboxyl latex gives the $\mathrm{pK_a}$ close to the literature value of carboxylic acids \cite{Borkovec2001}.
Fig.~\ref{fig:6} shows the titrated surface charge density of carboxyl latex colloids as a function of pH \cite{Borkovec2000}. 
The points are obtained by potentiometric titration, whereas the lines are Eq.~\ref{eq:Langmuir} with $\mathrm{pK_a}=4.9$ and $\Gamma_0=0.6\,$nm$^{-2}$ \cite{Borkovec2001}. 
The agreement between theory and experiments is quite good.  

\section{Novel effect on surface electrification}
\ \\
In the last section, the novel mechanisms of surface electrification are discussed in detail.
Other unique electrification mechanisms have been recognized, besides physical ion adsorption and chemical electrification.
For example, applying an external voltage to metallic electrodes in solutions is widely used in electrochemistry \cite{Grahame1947,Fenn2000}.
Gold colloids are normally charged because of ionic ligand coordination during synthesis \cite{Matej2020}. 
Charge transfer between water molecules at interfaces was also suggested as a mechanism of electrification  \cite{Vacha2012,Hassanali2020}. 
The electrification of clay and clay minerals is caused by isomorphic substitution for $\mathrm{Si}^{4+}$ by $\mathrm{Al}^{3+}$ and $\mathrm{Al}^{3+}$ by $\mathrm{Mg}^{2+}$, respectively, \cite{Tajana2016,Mugele2017}. 
In this section, we discuss two novel effects on surface electrification: flow effect \cite{Bonn2014, VanRoij2018, VanRoij2019, Bonn2020, Tanaka2018, Ram2018, Lutzen2018, Mcnamee2018, Mcnamee2019, Mcnamee2020} and contact electrification \cite{Zlich2008,Jarrold2011,Choi2013,Mesquida2013, Beauchamp2015, Mishra2020}.

\subsection{Flow effect}

\begin{figure}[t]
\includegraphics{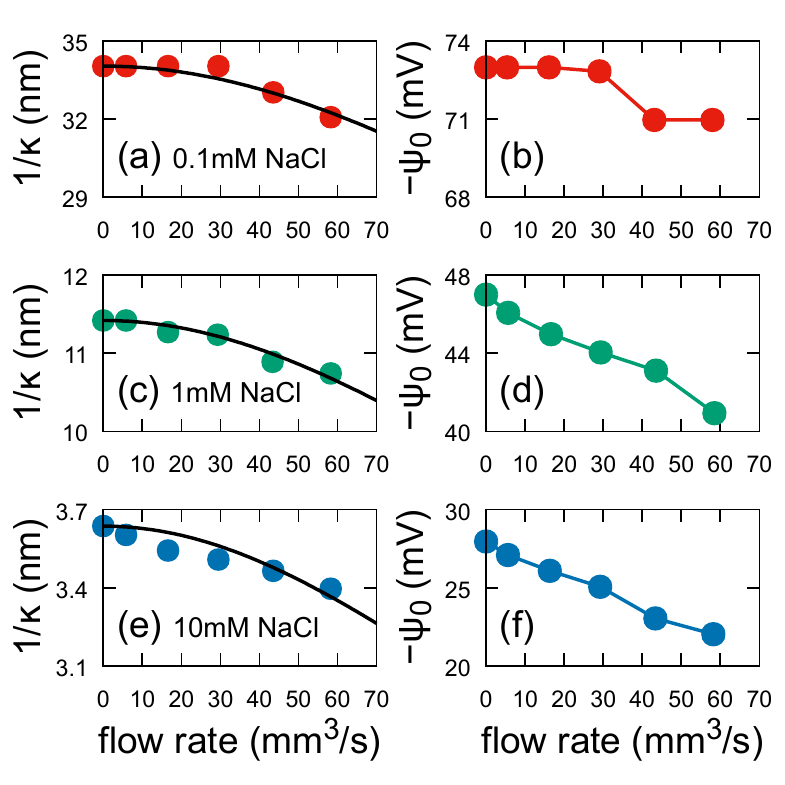}
\caption{The flow effect on the double-layer force between silica surfaces \cite{Mcnamee2018}.
The variation of the Debye length is plotted in (a,c,e), whereas the variation in the surface potential is plotted in (b,d,f).
The ionic strength was varied as (a,b) $0.1\,$mM, (c,d) $1\,$mM, and  (e,f) $10\,$mM.
The solid lines in (a,c,e) are the theoretical fits $\kappa=\kappa_0\sqrt{1+Cv^2}$ \cite{Ram2018}, where $v$ is the flow rate.
All the fit parameters are described in Appendix C.
The lines in (bdf) are guide for the eyes. 
}
\label{fig:14}
\end{figure}

It was reported that microfluidic channels composed of $\mathrm{CaF_2}$ or $\mathrm{SiO_2}$ exhibit  a reversible SFG spectrum response to the applied pulse flow \cite{Bonn2014}, suggesting that the surface charge of the channels depends on the flow. 
This flow-induced surface charge was understood by introducing the reaction dynamics of the dissociation on the surface with a very slow reaction time ($\sim 30\,$min) \cite{VanRoij2018, VanRoij2019}.  
A theoretical consideration of the ionic reaction dynamics suggested that the surface charge is inhomogeneous along the channel flow \cite{VanRoij2018, VanRoij2019}, and this was verified by position-resolved SFG spectroscopy \cite{Bonn2020}.

Another novel experimental study on the flow effect is the double-layer force measurement by atomic force microscopy in the presence of a liquid flow \cite{Mcnamee2018}.
Fig.~\ref{fig:14} shows the surface potential of the silica surface, and Debye length of the double layer repulsion forces as a function of the flow rate \cite{Mcnamee2018}.
A reduction in the surface potential and Debye length was observed for different salt concentrations.
The solid black lines in Fig.~\ref{fig:14} are theoretical predictions for a decrease in Debye length in a direction normal to the flow \cite{Ram2018}, which is described by the following equation: 
\begin{equation}
\frac{1}{\kappa} = \frac{1}{\kappa_0\sqrt{1+Cv^2}},
\label{eq:46}
\end{equation}
where $1/\kappa_0$ is the Debye length without flow, $v$ is the flow rate, and $C$ is the fitted parameter.
The agreement between theory \cite{Ram2018}, and experiments \cite{Mcnamee2018} reveals that the distortion of the electric double layer due to the external flow plays a crucial role in the reduction of the Debye length.
Further discussion of the fit parameters is described in Appendix C. 
These studies demonstrated  that ion reaction dynamics and external flows substantially affect the surface charge density under non-equilibrium conditions \cite{Bonn2020,Xi2020}.

\subsection{Contact electrification}

Contact electrification generally occurs when materials detach from different materials after contact \cite{Lacks2011,Wang2019}.  
In particular, this section focuses on the charging of water droplets during spray or pipetting.
Water droplets surrounded by an insulating medium, such as air or oil, are usually charged, and their charges strongly depend on the history of droplet formation \cite{Zlich2008,Jarrold2011, Beauchamp2015, Choi2013, Mesquida2013, Mishra2020, Burgo2020}.
Pipetting is the most common method to produce microdroplets.
The tips of pipettes are usually made of glass or polymeric materials, and are negatively charged.
Therefore, the droplets dispensed from these pipettes are usually positively charged \cite{Choi2013,Mesquida2013,Mishra2020}.
Electrospray is another example of droplet formation \cite{Colussi2006,Zlich2008,Beauchamp2015}, which is also used in mass spectrometry, as discussed in the previous section \cite{Colussi2006,Enami2010}. 

\begin{figure}[t]
\includegraphics{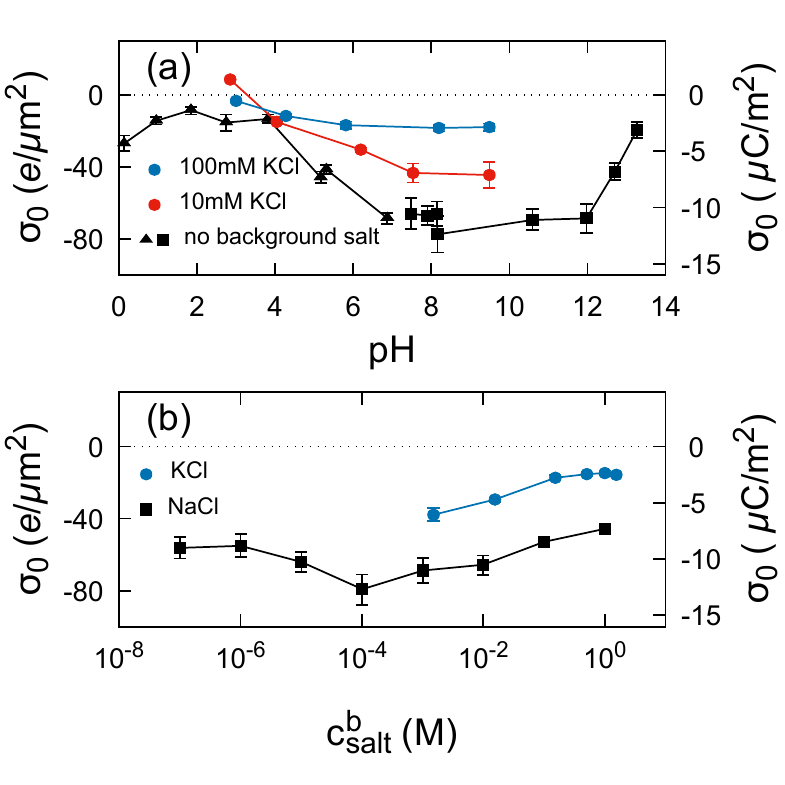}
\caption{
The surface charge density of a water/hydrophobic polymer interface is measured by charge separation during pipetting \cite{Choi2013,Mishra2020}. 
(a) pH dependence of the surface charge density obtained by pipetting droplets.
The black points \cite{Mishra2020} are data without background salts, whereas the colored points \cite{Choi2013} are data with 10 mM or 100 mM KCl salts.
(b) Salt concentration dependence of the surface charge density by pipetting droplets.
The black points \cite{Mishra2020} are data of NaCl, whereas the blue points are data of $7\,\mu$L droplets of KCl solution \cite{Choi2013}.
The tip of the pipette is made of polypropylene for black data \cite{Mishra2020}, whereas it is not specified for colored data \cite{Choi2013}.
The colored data in (a) and (b) were converted from the original charge amount by dividing the surface area by $32\,$mm$^2$ \cite{Choi2013}.
The sign of the charge was inverted from the original data to depict the surface charge of the tip surface.
}
\label{fig:16}
\end{figure}

Fig.~\ref{fig:16} shows the surface charge density of a polymeric tip of a pipette \cite{Choi2013,Mishra2020}.
The total amount of charge inside the droplets was measured by the Faraday cup method, and the surface charge density of the tip was obtained by dividing the total charge of the droplet by the inner surface area of the tip \cite{Choi2013,Mishra2020}.
The charging mechanism of the droplet is explained as follows.
During pipette loading, excess positive ions are carried into the water inside the tip because of the negative surface charge of the inner tip surface \cite{Mishra2020}.
After detachment of the tip from the water reservoir, the total charge of the pipette surface and the water inside the tip should be neutral.
However, because the negative surface charge is (partially) fixed on the tip surface, the droplet dispensed from the pipette is positively charged.  

Fig.~\ref{fig:16}a shows the pH dependence of the surface charge density of the pipette tip surfaces \cite{Choi2013,Mishra2020}. 
The qualitative behavior is very similar to the zeta potential measurements of hydrophobic surfaces (for Teflon AF, see Fig.~\ref{fig:1}) and the disjoining pressure measurements (see Fig.~\ref{fig:10} (b)]. 
The data for the $10\,$mM KCl solution (red circles) have an isoelectric point at $\mathrm{pH=3.5}$. 
The data without background electrolytes (black points) exhibit a negative charge under acidic conditions, in contrast to the zeta potential measurements.
This can be explained by the fact that the absorbed protons on the tip surfaces are transported to a dispensed droplet.
The dependence of the surface charge density on salt concentration is shown in Fig.~\ref{fig:16}b, suggesting that the tip surface behaves as a constant-charge surface rather than a constant-potential surface.
This method is unique to measure the surface charge density.
However, the surface charge density is much lower than that measured by the zeta potential measurement or potentiometric titration by a factor of $10^2$.

\section{Summary and perspective}
\ \\
In this review, various experimental methods to characterize the surface charge density of aqueous interfaces are discussed. 
Electrokinetic methods are most commonly used to measure the surface charge density, but it is inevitable to include the dynamic effect of the interface. 
On the other hand, potentiometric titration and double-layer force measurements provide a purely static property of the surface charge density. 
Surface tension measurement is another tool to analyze the surface affinity of ions. 

Surface-sensitive nonlinear spectroscopy and mass spectrometry are modern tools to analyze the interfacial structure of water interfaces. 
In particular, the theory for extracting the surface charge density from complex nonlinear polarizability is very useful for characterizing the charged surface.
Electrospray ionization mass spectrometry of electrolyte solution is also considered to be surface sensitive.
However,  the theoretical foundation for the surface sensitivity of ESI-MS is still under development.  

In addition to an overview of the experimental methods, the electrification mechanisms are discussed in detail.
Physical ion adsorption is a universal mechanism for creating surface charges for all types of surfaces.
In particular, the electrification of chemically inert interfaces, such as air/water interfaces, is completely governed by ion adsorption because the chemical electrification mechanism is lacking. 
The surface affinity of ions is strongly ion-specific, and the surface affinity of some ions, such as $\mathrm{H_3O}^+$ and $\mathrm{OH}^-$ to the air/water interface is still debated.
The impurity effect on surface electrification is discussed in detail because the significance of this effect has been recognized recently. 
Finally, proton dissociation of the carboxylic latex surface is explained in detail as an example of chemical electrification.

Recent intense research on charged water interfaces has revealed novel electrification mechanisms at water interfaces.
We take the flow effect and contact electrification as examples and review them in the final section.
It was experimentally demonstrated that the external flow along the surface modifies the surface charge density and Debye length. 
This can be explained by the ionic reaction dynamics at the surface and the hydrodynamics near the surface.
Another interesting effect is contact electrification occurring in electrospray and pipetting water droplets. 
This effect can be used as a unique method to measure the surface charge density of pipette tip surfaces. 

A variety of experimental methods to measure the surface charges have revealed quantitative and even qualitative mismatches of the surface charge densities of the same surface determined by different methods.
Making a continuous effort to resolve these contradicting mismatches provides further understanding of the structure and electrification mechanisms of charged interfaces.
We hope that this comprehensive review of the state-of-the-art progress of charged water interfaces will help to study and make a frontier in this field. 


\appendix

\section{Fit of the ionic potentials of mean force}

In this appendix, the fit functions for the potentials of the mean force plotted in Fig.~\ref{fig:12} are explained.
To fit the ionic potential of the mean force $U_i(z)$, we fit $\mathrm{e}^{- U(z)}$ by
\begin{equation}
 \mathrm{e}^{- U(z)} = \frac{1}{2}\left[1+\tanh\left( k(z-a)\right) \right],
\end{equation}
where $k$ and $a$ are fit parameters.
The above function does not work for $\mathrm{I^-}$, $\mathrm{H_3O^+}$, $\mathrm{Li}^+$, $\mathrm{Na^+}$, and $\mathrm{K^+}$.
Therefore, for these ions we use
\begin{equation}
 \mathrm{e}^{- U(z)} = \frac{1}{2}\left[1+\tanh\left( k(z-a)\right) \right] + n_0\mathrm{e}^{-p(z-b)^2},
\end{equation}
where $n_0$, $p$, and $b$ are the additional fit parameters.
The fitted parameters for each ion are tabulated in Table~\ref{tab:app} 

In a previous study \cite{Uematsu2018}, the surface affinity of sodium ions was calculated as $\alpha_\mathrm{Na}=1.16$ from the potential of the mean force of the force field parameter set 1 of  $\mathrm{Na}^+$ reported in Ref.~\citenum{Horinek2009}.
Using $\alpha_\mathrm{Cl}=0.98$ deduced from the potential of the mean force of Smith-Dang $\mathrm{Cl^-}$ \cite{Dang1994} at the air/water interface \cite{Horinek2009}, the theory predicts the experimental surface tension of the NaCl solution accurately \cite{Uematsu2018}.
Thus, both $\alpha_\mathrm{Na}=1.16$ and $\alpha_\mathrm{Cl}=0.98$ were determined by the potential of the mean force of sodium, and chloride ions in \cite{Uematsu2018}.

In this study, we use another fit function for the potential of mean force (or the ionic concentration profile), which has fewer parameters than those used previously \cite{Horinek2009}. 
In fact, this function fits the potential of the mean force of sodium ions at low energy, namely, $U_i(z)\lesssim 1$, better. 
This fit function gives $\alpha_\mathrm{Cl}=1.02$, which is almost equal to that in the previous study, 0.98, \cite{Uematsu2018}, whereas $\alpha_\mathrm{Na}=1.50$, which is larger than the previous one by 1.16 \cite{Uematsu2018}. 
Therefore, we reinterpret that $\alpha_\mathrm{Na}=1.2$ is obtained by fitting the experimental surface tension of the NaCl solution with the reference anion surface affinity $\alpha_\mathrm{Cl}=1.0$.
This reinterpretation allows us to consider that all the previous estimates of $\alpha_i^\mathrm{(exp)}$ other than $\mathrm{Na}^+$ and $\mathrm{Cl}^-$ from the fit with the experimental surface tension (tabulated in Table~\ref{tab:2}) \cite{Uematsu2018,Uematsu2019,Uematsu2018COLCOM,Uematsu2020} are still meaningful.

\begin{table}
\center
\caption{
The fit parameters of the potentials of mean force of ions.
}
\label{tab:app}
\vspace{2mm}
\begin{tabular}{cccccc}
Ion&$k$&$a$&$n_0$&$p$&$b$\\
&(nm$^{-1}$)&(nm)&&(nm$^{-2}$)&(nm)\\\hline
$\mathrm{F^-}$&5.02&0.425&---&---&---\\
$\mathrm{Cl^-}$&4.57&0.320&---&---&---\\
$\mathrm{Br^-}$&4.75&0.210&---&---&---\\
$\mathrm{I^-}$&9.82&0.045& \hspace{2.73mm}1.32&21.2&0.231\\
$\mathrm{Li^+}$&7.50&0.485&\hspace{2.73mm}0.69&18.6&0.279\\
$\mathrm{Na^+}$&6.81&0.467&\hspace{2.73mm}0.24&28.7&0.307\\
$\mathrm{K^+}$&2.82&0.264&$-0.19$&27.4&0.120\\
$\mathrm{Cs^+}$&5.49&0.315&---&---&---\\
$\mathrm{H_3O^+}$&3.63&0.268& \hspace{2.73mm}2.10&24.1&0.067\\
$\mathrm{OH^-}$&4.83&0.317&---&---&---\\
\end{tabular}
\end{table}

\section{$\alpha_i^\mathrm{(exp)}$ for iodide and lithium ions}

\begin{figure}[t]
\includegraphics{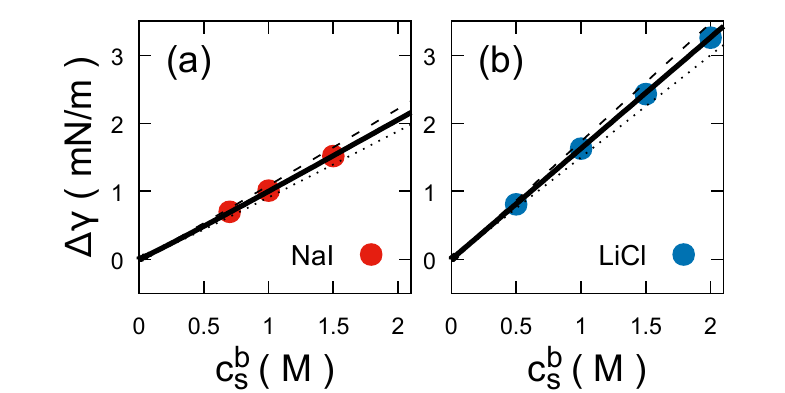}
\caption{
Surface tension of NaI and LiCl solutions.
The points are the experimental data \cite{Washburn1930}, whereas the line is calculated by the analytical interfacial layer model with $\varepsilon=78$, $T=298\,$K, $z^* = 0.5\,$nm \cite{Uematsu2018}.
(a) $\alpha_\mathrm{I} = 0.0$, and $\alpha_\mathrm{Na}=1.2$. $\alpha_\mathrm{I}=0.0\pm 0.1$ is denoted by broken and dotted lines.
(b) $\alpha_\mathrm{Li} = 1.2$, and $\alpha_\mathrm{Cl}=1.0$. $\alpha_\mathrm{Li}=1.2\pm 0.3$ is denoted by broken and dotted lines.
}
\label{fig:13}
\end{figure}

In this appendix, we extract the surface affinities of iodide and lithium ions at the air/water interface from the experimental surface tension data \cite{Washburn1930}.
The surface affinities of these ions have not been obtained by fitting the analytical interfacial layer model \cite{Uematsu2018}.
Therefore, we fit $\alpha_\mathrm{I}$ and $\alpha_\mathrm{Li}$ by fixing $\alpha_\mathrm{Na}=1.2$ and $\alpha_\mathrm{Cl}=1.0$, as shown in Fig.~\ref{fig:13}.
The obtained surface affinities are $\alpha_\mathrm{I}=0.0$ and $\alpha_\mathrm{Li}=1.2$, which are tabulated in the column of  $\alpha_i^\mathrm{(exp)}$ in Table~\ref{tab:2}.

\section{Fit of the reduction of Debye length by flow effect}

\begin{table}
\center
\caption{
The fit parameter $C$ of the reduction in the Debye length by the flow effect.
}
\label{tab:app2}
\vspace{2mm}
\begin{tabular}{ccc}
$c_\mathrm{s}\,$(mM)&$1/\kappa_0\,$(nm)&$C\,$(s/mm$^3$)$^2$\\\hline
0.1&34&$3.4\times 10^{-5}$\\
1&11&$4.2\times 10^{-5}$\\
10&3.6&$5.0\times 10^{-5}$\\
\end{tabular}
\end{table}

In this appendix, the details of fitting  Eq.~\ref{eq:46} in Fig.~\ref{fig:14}ace is explained.
The fit parameter $C$ is tabulated in Table~\ref{tab:app2}.
The equilibrium Debye length, $1/\kappa_0$, was extracted from the experimental data without flow rate, and it was fixed during the fit.
Assuming the flow rate $Q=L^2 u$, where $L$ is the characteristic length of the channel and $u$ is the flow velocity, the analytical theory predicts $C=(4D_\mathrm{ion}\kappa_0 L^2)^{-2}\sim (1/\kappa_0)^2$, where $D_\mathrm{ion}$ is the diffusion constant of the ion \cite{Ram2018}.
However, the fitted $C$ exhibits $C\sim\textrm{const}$ for different Debye lengths.  
This is because of the special geometry of the experimental channel \cite{Mcnamee2018}.
The fitted $C$ suggests that $Q=L u/\kappa_0$ instead of $Q=L^2 u$. 
 
\section*{References} 
\bibliographystyle{iopart-num}
\bibliography{paper_jpcm}

\end{document}